\documentclass[nofootinbib,aps,11pt]{revtex4-1}
\usepackage{graphicx}
\usepackage{hyperref}
\usepackage{cancel}
\usepackage{amssymb}
\usepackage{textcomp}
\usepackage{amsmath}
\usepackage{bm}
\usepackage{times}
\usepackage{epsfig}
\usepackage{color}

\newcommand{\eV}{{\rm eV}}
\newcommand{\MeV}{{\rm MeV}}
\newcommand{\GeV}{{\rm GeV}}
\newcommand{\TeV}{{\rm TeV}}
\newcommand{\fb}{{\rm fb}}
\newcommand{\pb}{{\rm pb}}

\begin{document}
\title{\LARGE  $Z'$ Portal Dark Matter in $B-L$ Scotogenic Dirac Model}
\bigskip
\author{Zhi-Long Han$^1$}
\email{sps\_hanzl@ujn.edu.cn}
\author{Weijian Wang$^2$}
\email{wjnwang96@aliyun.com}
\affiliation{
$^1$School of Physics and Technology, University of Jinan, Jinan, Shandong 250022, China
\\
$^2$Department of Physics, North China Electric Power University, Baoding 071003,China}

\date{\today}

\begin{abstract}
In this paper, we perform a detail analysis on the phenomenology of $Z'$ portal scalar and Dirac fermion dark matter in $B-L$ scotogenic Dirac model. Unconventional $B-L$ charge $Q$ is assigned to the right-handed neutrino $\nu_R$ in order to realise scotogenic Dirac neutrino mass at one-loop level, where three typical value $Q=-\frac{1}{4},-4,\frac{3}{2}$ are chosen to illustrate. Observational properties involving dilepton signature at LHC, relativistic degrees of freedom $N_\text{eff}$, dark matter relic density, direct and indirect detections are comprehensively studied. Combined results of these observables for the benchmark scenarios imply that the resonance region $M_\text{DM}\sim M_{Z'}/2$ is the viable parameter space. Focusing on the resonance region, a scanning for TeV-scale dark matter is also performed to obtain current allowed and future prospective parameter space.
\end{abstract}

\maketitle

\section{Introduction}

Despite the unambiguous confirmation of non-zero neutrino masses and mixings, the nature of neutrinos, which could be Majorana or Dirac, is still an open question. It is usually argued that tiny neutrino masses may naturally originate from the effective $\Delta L=2$ lepton number violating (LNV) operators\cite{Weinberg:1979sa}, so the Majorana scenario seems more promising.  However, the neutrinoless double beta decay ($0\nu 2\beta$) experiments, aimed to search for $\Delta L=2$ LNV signatures, have been performed without any positive result so far and the possibility that neutrinos are Dirac particles can not be excluded. In the standard model (SM) along with three copies of $\nu_{R}$, Dirac neutrinos can acquire their masses via direct
Yukawa coupling $y_{\nu}\overline{L}H\nu_{R}$. In order to generate sub-eV neutrino masses, the coupling constants $y_{\nu}$ have to be fine tuned to $10^{-12}$ order. Consequently, if neutrinos are Dirac particles, certain new physics beyond SM should exist to explain the extremely small Yukawa coupling constants in a natural way. Motivated by these considerations, interest in Dirac neutrino masses has been revived recently and several models\cite{Gu:2006dc, Gu:2007ug, Farzan:2012sa, Chulia:2016ngi,Bonilla:2016diq,Ma:2016mwh,Wang:2016lve,Borah:2017leo,Wang:2017mcy,CentellesChulia:2017koy, Ma:2017kgb,Yao:2017vtm,Bonilla:2017ekt,Ibarra:2017tju,Borah:2017dmk,Yao:2018ekp,CentellesChulia:2018gwr,
CentellesChulia:2018bkz} are proposed at tree or loop level.

On the other hand, the connections between LNV effects and the nature of neutrinos are often overlooked. For example, Dirac neutrinos are compatible with standard baryogenesis processes\cite{tHooft:1976rip,Kuzmin:1985mm}, where the non-perturbative effect breaks the baryon ($B$) and lepton ($L$) number each by three units while $B-L$ number remains conserved.
As is well known, the simplest anomaly-free $B-L$ extension of SM includes three copies of $\nu_{R}$ having $-1$ charge under $U(1)_{B-L}$ symmetry. Then the Majorana neutrino will not arise if $U(1)_{B-L}$ symmetry never breaks down by two units, i.e., $\Delta (B-L)\neq 2$, which explains why neutrinos are Dirac fermions. In Ref. \cite{Heeck:2013rpa, Heeck:2013vha}, the quartic LNV operators i.e., $\Delta (B-L)= 4$ processes are considered as new sources for leptogenesis even if neutrinos are Dirac particles.

However, the simple $B-L$ extension of SM model can not explain the extremely small Yukawa coupling constant in the presence of $y_{\nu}\overline{L}H\nu_{R}$ term.  It is reasonable to ask if gauged $U(1)_{B-L}$ symmetry can play a key role in the generation of naturally small Dirac neutrino masses. Recently, a class of $B-L$ scotogenic models are proposed by us in Ref.\cite{Wang:2017mcy} where the Dirac neutrino masses are generated at loop level. The Greek ``scotogenic" means darkness. The original model, firstly propose by Ma \cite{Ma:2006km}, is to accommodate two important missing pieces of SM: no-zero Majorana neutrino mass and dark matter (DM) in a unified framework. The main idea is based on the assumption that the DM candidates, whose the stability is protected by an $\emph{ad hoc}$ $Z_{2}$ or $Z_{3}$ \cite{Ding:2016wbd} symmetry, can serve as intermediate messengers propagating inside the loop diagram in neutrino mass generation. In the scotogenic Dirac  model, the $B-L$ quantum number of $\nu_{R}$ are appropriately assigned so that the $y_{\nu}\overline{L}H\nu_{R}$ interaction and Majorana term $\overline{\nu}_{R}\nu_{R}^{c}$ are forbidden. Some new Dirac fermions are introduced to be intermediate particles carrying $B-L$ quantum numbers. Since all new particles are SM singlets, we only need to consider the $[U(1)_{B-L}]\times[Gravity]^{2}$ and $[U(1)_{B-L}]^{3}$ anomaly free conditions.
Then the effective $\overline{\nu_{L}}\nu_{R}$ operator is induced after the spontaneous breaking of $U(1)_{B-L}$ at high scale. Moreover,  the discrete $Z_{2}$ or $Z_{3}$ symmetry could appear as a remnant symmetry of gauged $U(1)_{B-L}$ symmetry, leading to DM candidates.

In previous studies \cite{Wang:2017mcy}, we focus on scalar interactions of DM in $B-L$ scotogenic Dirac model. As complementary, this paper will concentrate on gauge interactions of DM.
Early researches on various $Z'$ portal DM can be found in Refs.~\cite{Alves:2013tqa,
Buchmueller:2014yoa,Alves:2015pea,Wang:2015saa,Alves:2015mua,Jacques:2016dqz,Fairbairn:2016iuf,
Kaneta:2016vkq,Altmannshofer:2016jzy,Singirala:2017see,Ding:2017jdr,Han:2017ars,Ma:2018bjw,Okada:2018ktp,
FileviezPerez:2018toq,Biswas:2018yus}.
As for Majorana fermion DM in conventional $B-L$ scotogenic Majorana model for neutrino masses~\cite{Kanemura:2011vm}, no observable signatures are expected at incoming direct or indirect detection experiments except for LHC dilepton signature of $Z'$. In contrast, the $Z'$ portal scalar DM in $B-L$ scotogenic Dirac model could be tested at direct detection experiments, and Dirac fermion DM could be further probed by indirect detection. Therefore, if the ongoing DM experiments observe some positive signatures, the nature of DM might be revealed.

Moreover, the existence of right-handed neutrino $\nu_R$ will also contribute to effective number of relativistic degrees of freedom  $N_\text{eff}$ for neutrinos via $Z'$ portal, which can also be used to verify this model. Note that although the well studied simplified $Z'$ portal DM approach is constructive, there might be some different predictions for a complete model, especially when a new observable as $N_\text{eff}$ in this model is involved. So in this paper, we perform a comprehensive analysis on the phenomenology of $Z'$ portal dark matter in $B-L$ scotogenic Dirac model in aspect of collider signatures, dark radiation, relic density, direct and indirect detection.

The rest of the paper is organised as following. In Sec~\ref{Sec:MD}, we briefly review the minimal gauged $U(1)_{B-L}$ scotogenic Dirac model. Collider signatures from LHC are discussed in Sec~\ref{Sec:LHC}. The exclusion limits are derived from the latest dilepton signature search. Dark radiation observable $N_\text{eff}$ and corresponding limits are then considered in Sec.~\ref{Sec:DR}. The DM phenomenology in aspect of relic density, direct detection and indirect detection are detail studied in Sec.~\ref{Sec:DM} for some benchmark scenarios. In Sec.~\ref{Sec:CA}, we first present the combined results for certain benchmark scenarios. And then based on the combined results, we perform a scan over the resonance region $M_\text{DM}\simeq M_{Z'}/2$. Finally, the conclusion is devoted to Sec.~\ref{Sec:CL}.

\section{The Model}\label{Sec:MD}

\begin{table}[!htbp]
\begin{tabular}{|c|c|c|c|c|c|c|c|c|}
  \hline
  Particles & ~$L$~ & ~$\nu_R$~ & ~$F_L$~ & ~$F_R$~ & ~$\Phi$~ & ~$\eta$~ & ~$\chi$~ & ~$\sigma$~
  \\ \hline
  $SU(2)_L$ & $2$ & $1$ & $1$ & $1$ & $2$ & $2$ & $1$ &  $1$
  \\ \hline
  $U(1)_Y$ & $-\frac{1}{2}$ & $0$ & $0$ & $0$ & $\frac{1}{2}$ & $\frac{1}{2}$ & $0$ & $0$
  \\ \hline
  $U(1)_{B-L}$ & $-1$ & $Q$ & $1$ & $-Q$ & $0$ & $Q-1$ & $Q-1$ & $Q+1$
  \\ \hline
\end{tabular}
\caption{Particle contents and corresponding charge assignment.}
\label{TB:contents}
\end{table}

The gauged $U(1)_{B-L}$ scotogenic Dirac neutrino mass models are systematically studied in Ref.~\cite{Wang:2017mcy}. In this paper, we concentrate on the minimal $U(1)_{B-L}$ scotogenic Dirac model, i.e., model $A_1$ in Ref.~\cite{Wang:2017mcy}. The relevant particle contents and corresponding charge assignment are explicitly shown in Table~\ref{TB:contents}. In addition to three generations of right-handed neutrino $\nu_R$ with $B-L$ charge $Q$, this model further employs three generations of Dirac fermion $F$ with $B-L$ charge $+1$ for the left-handed component and $-Q$ for the right-handed component.  In the scalar sector, this model introduces one scalar doublet $\eta$, two scalar singlets $\chi$ and $\sigma$ with $B-L$ charge $Q-1$, $Q-1$ and $Q+1$ respectively.
Owing to newly introduced chiral fermions being singlets under the SM gauge group, one only need to check the $[U(1)_{B-L}]\times[Gravity]^2$ and $[U(1)_{B-L}]^3$ anomaly free conditions \cite{Ma:2001kg}
\begin{eqnarray}
-3-3Q~~-3(-Q)~~+3\times1&=&0,\\
-3-3Q^3-3(-Q)^3+3\times1&=&0.
\end{eqnarray}
Obviously, the anomalies has been canceled for per generation fermions, since $F_{L,R}$ has opposite $B-L$ charge comparing with $\nu_{L,R}$. Therefore, $Q$ can be regarded as a free parameters apart from the following exceptions:
\begin{itemize}
  \item To forbid the SM direct Yukawa coupling term $\bar{\nu}_L\nu_R\overline{\phi^0}$, the condition $Q\neq-1$ should be imposed.
  \item To forbid Majorana mass terms
  $(m_{R})\overline{\nu_{R}^C}\nu_{R}$, $\sigma\overline{\nu_{R}^C}\nu_{R}$ and $\sigma^{\ast}\overline{\nu_{R}^C}\nu_{R}$, the conditions $Q\neq0, -1/3$ and $ 1$ should also be satisfied respectively.
  \item To avoid tree-level neutrino mass, linear terms as $\sigma^{k}\chi$ and $(\sigma^{\ast})^{k}\chi$($k=1,2,3$) must be forbidden, which then requires $Q\neq 0, -1/3,-1/2,-2$ and $-3$.
  \item Similarly, forbidding the $(\Phi^{\dag}\eta)\sigma^{k}$ and $(\Phi^{\dag}\eta)(\sigma^{\ast})^{k} (k=1,2)$ terms leads to $Q\neq 0, -1/3,-3$.
\end{itemize}
Once an appropriate $Q$ is assigned, the residual $Z_{2}$ symmetry appears in Eq.~\ref{Eq:LZp}, \ref{Eq:pot} and \ref{Eq:Yuk}, under which the parity is odd for inert particles ($\eta,\chi,F_{L/R}$) and even for all other particles.
In the following, we consider $Q=-\frac{1}{4},-4$ and $\frac{3}{2}$ for illustration.

The relevant new gauge interactions are dictated by
\begin{eqnarray}\label{Eq:LZp}
\mathcal{L}_{Z'} \supset && -\frac{1}{4} F'^{\mu\nu} F'_{\mu\nu} + (D^\mu \eta)^\dag(D_\mu \eta) + |D^\mu \chi|^2 + |D^\mu \sigma|^2 + \frac{g'}{3} \sum_{i=1}^6( \bar{q}_i\gamma^\mu q_i) Z'_\mu
\\ \nonumber
 &&  +
 g' \sum_{i=1}^3(- \bar{\ell}_i \gamma^\mu \ell_i - \bar{\nu}_{Li}\gamma^\mu \nu_{Li}
 + Q \bar{\nu}_{Ri}\gamma^\mu \nu_{Ri} + \bar{F}_{Li}\gamma^\mu F_{Li}
 - Q \bar{F}_{Ri}\gamma^\mu F_{Ri}) Z'_\mu,
\end{eqnarray}
where $D_\mu=\partial_\mu-ig'Q_{BL}Z'_\mu$.
The complete scalar potential is given by \cite{Wang:2017mcy}
\begin{eqnarray}\label{Eq:pot}
V&=& -\mu_\Phi^2 \Phi^\dag\Phi + \mu_\eta^2 \eta^\dag\eta + \mu_\chi^2 \chi^*\chi - \mu_\sigma^2 \sigma^*\sigma +\lambda_\Phi^2 (\Phi^\dag\Phi)^2+ \lambda_\eta (\eta^\dag \eta)^2 + \lambda_\chi (\chi^*\chi)^2 \\ \nonumber
&& +\! \lambda_\sigma (\sigma^*\sigma)^2\! +\! \lambda_{\Phi \eta} (\Phi^\dag\Phi) (\eta^\dag \eta)\! +\!  \lambda'_{\Phi \eta} (\eta^\dag\Phi) (\Phi^\dag \eta) \! +\!  \lambda_{\Phi \chi} (\Phi^\dag\Phi) (\chi^*\chi) \! +\!  \lambda_{\Phi \sigma} (\Phi^\dag\Phi) (\sigma^*\sigma)
\\ \nonumber
&& + \lambda_{\eta \chi} (\eta^\dag \eta) (\chi^*\chi) + \lambda_{\eta \sigma} (\eta^\dag \eta) (\sigma^*\sigma) + \lambda_{\chi\sigma} (\chi^*\chi) (\sigma^*\sigma) +
\left[\mu(\Phi^{\dagger}\eta)\chi^{\ast} + \text{h.c.}\right],
\end{eqnarray}
where $\mu^2_X(X=\Phi,\eta,\chi,\sigma)$ are all taken to be positive, and $\mu$ is also positive by proper re-phasing $\eta$ and $\chi$. Due to the lack of $(\Phi^\dag \eta)^2$ term, masses of $\eta_R^0$ and $\eta_I^0$ are degenerate as a complex scalar $\eta^0$. Because $\eta^0$ has direct coupling with $Z$, only the singlet $\chi$ is suitable for scalar DM. To concentrate on $Z'$-portal DM in this paper, the Higgs-portal $\lambda_{\Phi \chi}$, $\lambda_{\Phi\sigma}$ and $\lambda_{\chi\sigma}$ terms are assumed to be tiny.

\begin{figure}
\begin{center}
\includegraphics[width=0.45\linewidth]{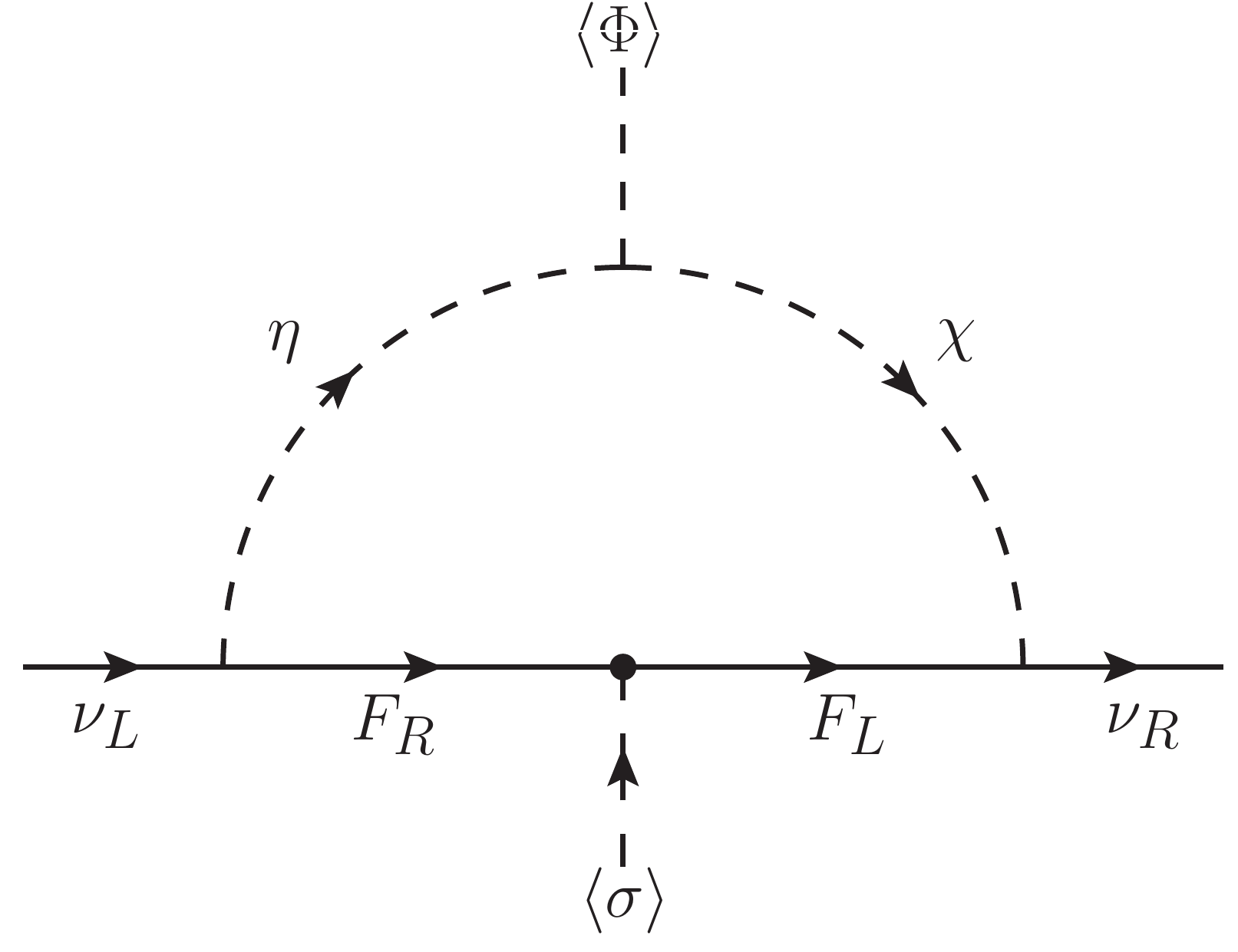}
\end{center}
\caption{One-loop generation of Dirac neutrino mass.
\label{Fig:Rv}}
\end{figure}

After spontaneous symmetry breaking, the scalar $\Phi$ and $\sigma$ are denoted as
\begin{align}
\Phi=\left(
\begin{array}{c}
0\\
\frac{v_{\phi}+\phi^0}{\sqrt{2}}
\end{array}\right),
\quad \sigma = \frac{v_{\sigma}+ \varphi^0}{\sqrt{2}},
\end{align}
in unitarity gauge. Here, $v_\phi=246~\GeV$ is the electroweak scale. Then the $\lambda_{\Phi\sigma}$ term induces mixing between $\phi^0$ and $\varphi^0$ with mixing angle $\alpha$, resulting to mass eigenstates $h^0$ and $H^0$ \cite{Ding:2018jdk}. In this paper, we regard $h^0$ as the $125~\GeV$ Higgs boson discovered at LHC \cite{Aad:2012tfa,Chatrchyan:2012xdj}, and $H^0$ is a heavier scalar singlet \cite{Robens:2015gla}. Meanwhile, the $Z_2$-odd scalar $\chi$ and $\eta^0$ mix into $H_1^0$ and $H_2^0$ with mixing angle $\beta$. In this paper, we assume $M_{H_1^0}<M_{H_2^0}$ and $\beta\ll 1$, so $H_1^0$ is dominant by $\chi$-component and is regarded as scalar DM candidate. Besides, the $Z_2$-odd charged scalar $H_2^\pm(=\eta^\pm)$ do not mix with other particles.

The Yukawa interactions accounting for radiative Dirac neutrino mass are given as \cite{Wang:2017mcy}
\begin{equation}\label{Eq:Yuk}
\mathcal{L}_Y = y_{1}\overline{L}F_{R}i\tau_{2}\eta^{\ast}+
y_{2}\overline{\nu_{R}}F_{L}\chi+ y_F\overline{F_{L}}F_{R}\sigma+\text{h.c.}.
\end{equation}
Hence, VEV of $\sigma$ will induce Dirac fermion mass $M_F= y_F v_\sigma/\sqrt{2}$, the lightest of which, i.e., $F_1$ serves as fermion DM candidate. As shown in Fig.~\ref{Fig:Rv}, the above Yukawa interactions will induce Dirac neutrino mass at one-loop level. The neutrino mass matrix can be expressed as
\begin{equation}\label{Eq:mv}
m_{\nu}^{ij}= \frac{\sin2\beta}{32\pi^2}\sum_k y_1^{ik} y_2^{jk*}
M_{F_k}\left[\frac{M_{H_2^0}^2}{M_{H_2^0}^2-M_{F_k}^2}\ln\left(\frac{M_{H_2^0}^2}{M_{F_k}^2}\right)
-\frac{M_{H_1^0}^2}{M_{H_1^0}^2-M_{F_k}^2}\ln\left(\frac{M_{H_1^0}^2}{M_{F_k}^2}\right)\right].
\end{equation}
To accommodate neutrino mass around $0.1~\eV$, $\beta\sim10^{-4}$, $y_{1,2}\sim0.01$ with masses of $Z_2$-odd particles at TeV-scale are required.

In the following, we concentrate on the phenomenology of $Z'$ portal DM. In this case, there are only four free parameters: the new gauge coupling $g'$, the new gauge boson mass $M_{Z'}$, the DM mass $M_\text{DM}$ and $\nu_R$'s $B-L$ charge $Q$. Before investigate the DM phenomenon, we consider possible constraints from collider signature in Sec.~\ref{Sec:LHC} and dark radiation in Sec.~\ref{Sec:DR}.

\section{Collider Signature}\label{Sec:LHC}

\begin{table}
\begin{tabular}{|c|c|c|c|}
  \hline
  $Q$-value & $q\bar{q}$ & $\ell^+ \ell^-$ & $\nu\bar{\nu}$
  \\ \hline
  $Q=-\frac{1}{4}$ / $-4$ / $\frac{3}{2}$ & ~0.30 / 0.07 / 0.20 ~ & ~0.45 / 0.10 / 0.30~ & ~0.25 / 0.83 / 0.50~
  \\ \hline
\end{tabular}
\caption{Branching ratio of $Z'$ for different value of $Q$.}
\label{TB:BRZp}
\end{table}

Spontaneous breaking of the $U(1)_{B-L}$ gauge symmetry by  the VEV of $\sigma$ induces the mass term of the $U(1)_{B-L}$ gauge boson as $M_{Z'}=Q_\sigma g' v_\sigma=(Q+1)g' v_\sigma$. The precise measurement of four-fermion interactions at LEP requires\cite{Cacciapaglia:2006pk}
\begin{equation}\label{Eq:LEP}
\frac{M_{Z'}}{g'}\gtrsim7~\TeV.
\end{equation}
In the limit that masses of SM fermions
$f$($f\equiv q, l, \nu$) are small compared with the
$Z^{\prime}$ mass, the decay width of $Z^{\prime}$ into
fermion pair $f\overline{f}$ is given by
\begin{equation}
\Gamma(Z^{\prime}\rightarrow
f\overline{f})=\frac{g'^{2}M_{Z^{\prime}}}{24\pi}N_C^{f}(Q_{fL}^{2}+Q_{fR}^{2})
\end{equation}
where $N_C^f$ is the number of colours of the fermion $f$, i.e., $N_C^{l,\nu}=1$, $N_C^{q}=3$. Then the branch ratios of $Z^{\prime}$ decay into each final states take the ratios as
\begin{equation}\label{Eq:BRZp}
\text{BR}(Z^{\prime}\rightarrow q\overline{q}):\text{BR}(Z^{\prime}\rightarrow
l^{+}l^{-}):\text{BR}(Z^{\prime}\rightarrow
\nu\overline{\nu})=4:6:3(1+Q^{2}).
\end{equation}
The $Q$-value has a great impact on the decay properties of $Z'$. According to Eq.~\ref{Eq:BRZp}, the dominant decay mode of $Z'$ is $Z'\to\ell^+\ell^-$ for $|Q|<1$, and $Z'\to \nu\bar{\nu}$ for $|Q|>1$.
A viable pathway to verify the $Q$-value is via the measurement of the ratio $\text{BR}(Z'\to\nu\bar{\nu})/\text{BR}(Z'\to\mu^+\mu^-)=3(1+Q^2)/2$.
In Table~\ref{TB:BRZp}, we depict the explicit value of branching ratios of $Z'$ for $Q=-\frac{1}{4}$, $-4$ and $\frac{3}{2}$ respectively. For $Q=-\frac{1}{4}$, the dominant decay mode $Z'\to \ell^+\ell^-$ has a branching ratio of $0.45$. In comparison,  the branching ratio of this dilepton mode diminishes to 0.10 when $Q=-4$, and the invisible decay mode $Z'\to \nu\bar{\nu}$ with a branching ratio 0.83 becomes the dominant one.

The most promising signature of $U(1)_{B-L}$ gauge boson at LHC is the dilepton signature $pp\to Z' \to \ell^+\ell^-(\ell=e,\mu)$ \cite{Basso:2008iv}. Searches for the new gauge boson $Z'$ in the dilepton final states have been performed by ATLAS \cite{Aad:2014cka} and CMS \cite{Khachatryan:2016zqb}. Due to no observation of any excess, the most stringent upper limit on the cross-section times branching ratio ($\sigma\text{BR}$) has been set by ATLAS using $36.1~\fb^{-1}$ data at $\sqrt{s}=13~\TeV$ LHC \cite{Aaboud:2017buh}. Here, this upper limit will be interpreted in the the frame work of $B-L$ scotogenic Dirac model. See Ref.\cite{Okada:2016gsh,Klasen:2016qux,Okada:2016tci,DeRomeri:2017oxa} for some earlier studies.

\begin{figure}
\begin{center}
\includegraphics[width=0.45\linewidth]{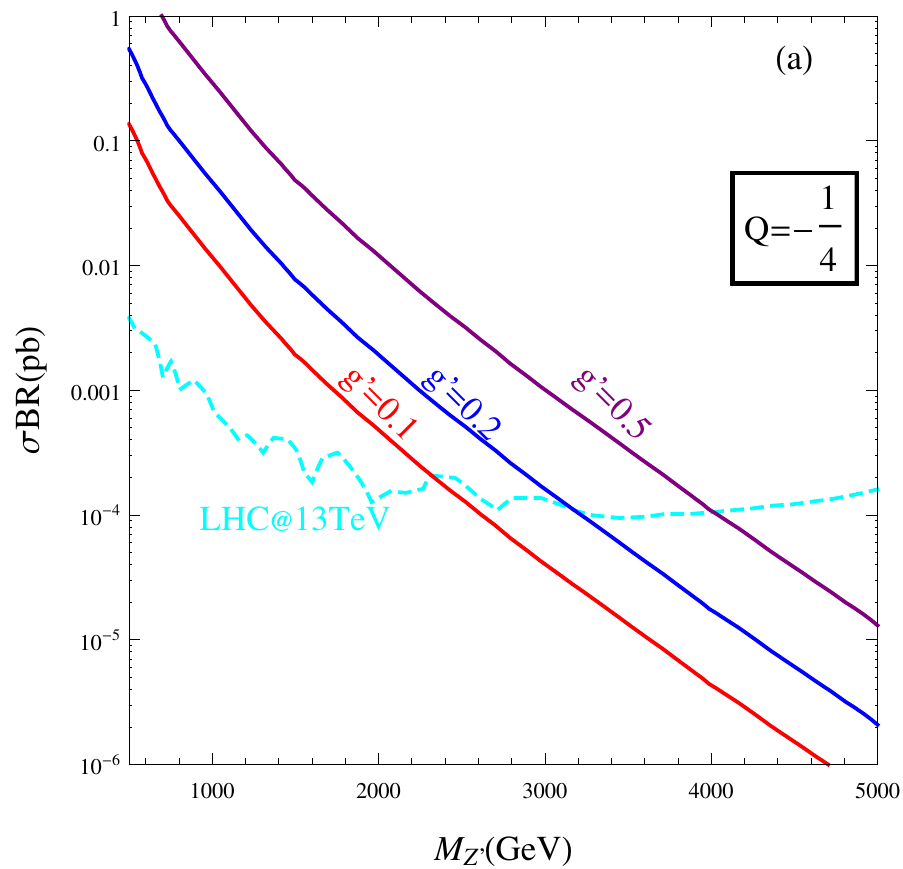}
\includegraphics[width=0.45\linewidth]{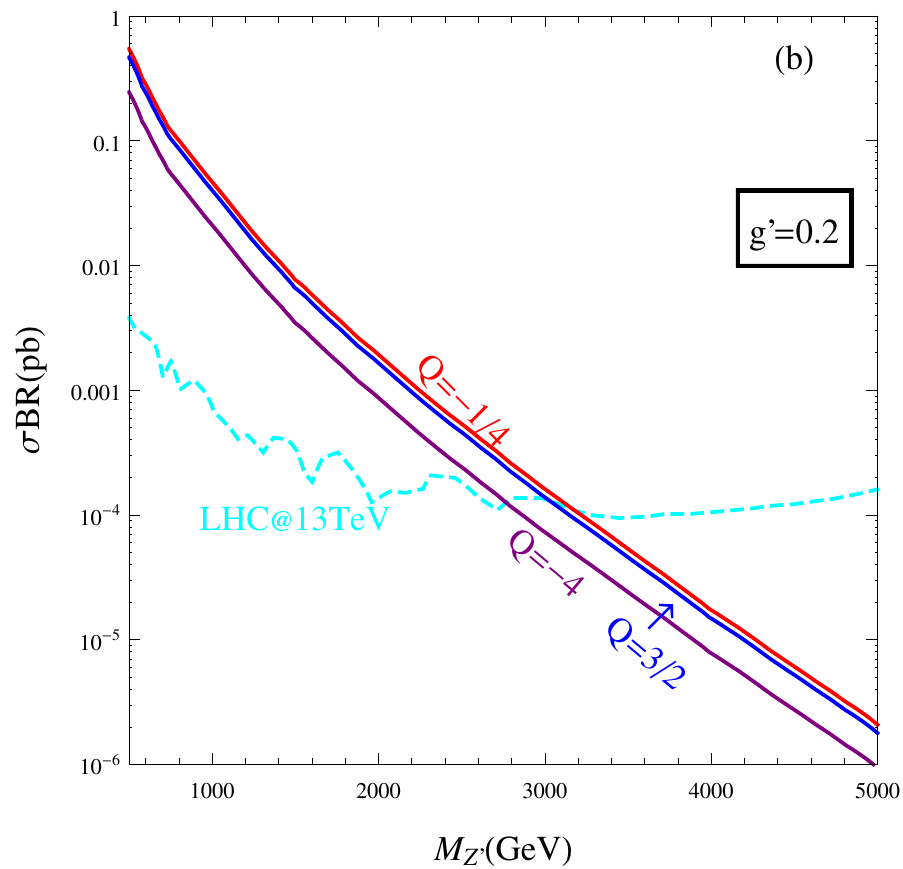}
\end{center}
\caption{Cross section of the dilepton signature $pp\to Z'\to \ell^+\ell^-$ at $13~\TeV$ LHC. (a) $Q=-\frac{1}{4}$, $g'=0.1,0.2,0.5$. (b) $g'=0.2$, $Q=-\frac{1}{4}, -4, \frac{3}{2}$. The dashed line is the upper limit from Ref.~\cite{Aaboud:2017buh}.
\label{Fig:LHCcs}}
\end{figure}

In carrying out the theoretical cross section of the dilepton signature, we implement the scotogenic Dirac model into {\tt FeynRules} \cite{Alloul:2013bka} and employ {\tt MadGraph5\_aMC@NLO} \cite{Alwall:2014hca} with {\tt NNPDF} \cite{Ball:2014uwa} parton distribution function. As suggested by Ref.~\cite{Okada:2016gsh}, a factor of $k=1.20$ is also multiplied in order to include the QCD corrections. The results are shown in Fig.~\ref{Fig:LHCcs}. Clearly, for a fixed $Q$-value, a larger $g'$ leads to a larger cross section (shown in Fig.~\ref{Fig:LHCcs} (a)). Meanwhile, for a fixed $g'$-value, a larger $|Q|$ leads to a smaller cross section (shown in Fig.~\ref{Fig:LHCcs} (b)), mainly due to the suppression of $\text{BR}(Z'\to \ell^+\ell^-)$.

\begin{figure}
\begin{center}
\includegraphics[width=0.45\linewidth]{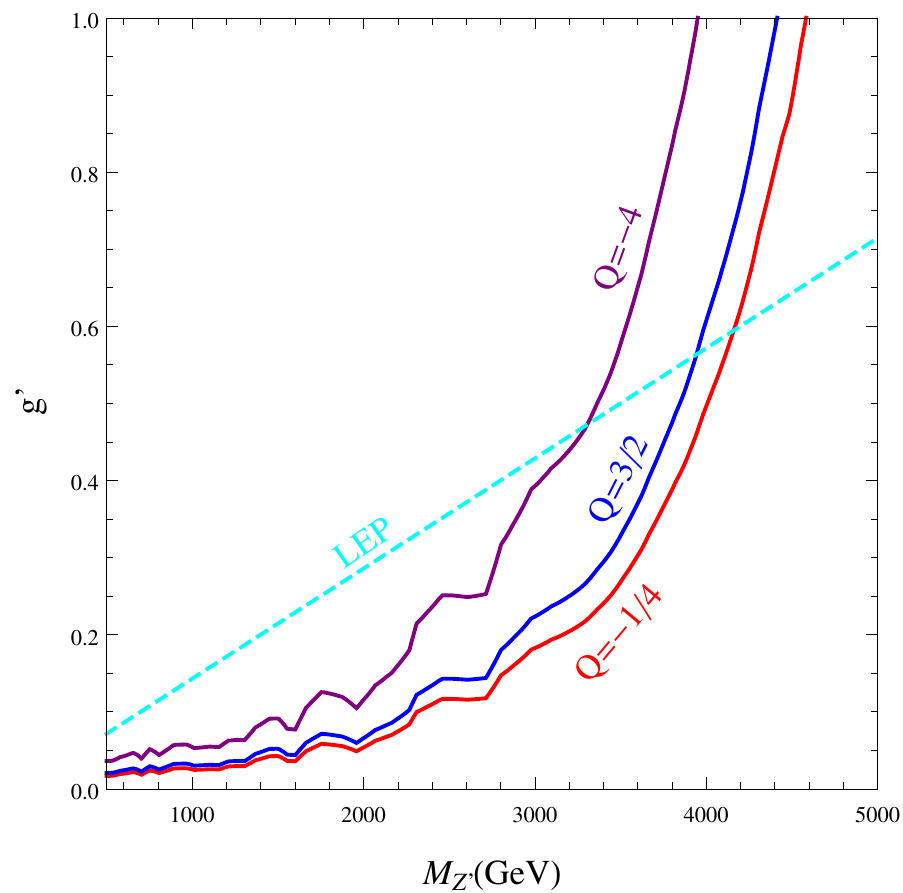}
\end{center}
\caption{LHC exclusion limits in the $g'-M_{Z'}$ for $Q=-\frac{1}{4},-4,\frac{3}{2}$ in the $U(1)_{B-L}$ scotogenic Dirac model. The dashed line corresponds to the LEP limit in Eq.~\ref{Eq:LEP}.
\label{Fig:LHCex}}
\end{figure}

By comparing the theoretical prediction and experimental limit, one can easily obtain the exclusion limit in the $g'-M_{Z'}$ plane, which is depicted in Fig.~\ref{Fig:LHCex} for $Q=-\frac{1}{4},-4,\frac{3}{2}$ respectively. Basically speaking, the exclusion limit for $Q=-4$ are the weakest, while exclusion limits are similar for $Q=-\frac{1}{4}$ and $Q=\frac{3}{2}$. Taking $Q=-\frac{1}{4}$ for an instance, the lower limit of $Z'$ mass is about $4~\TeV$ with $\mathcal{O}(1)$ $g'$-value, and this lower limit could down to about $3~\TeV$ with $g'\sim0.1$. In the lower mass region $M_{Z'}\lesssim 3~\TeV$, the limit from LHC dilepton signature is more stringent than LEP. But for $M_{Z'}\gtrsim4~\TeV$, the LEP limit becomes severer than the LHC limit.

\section{Dark Radiation}\label{Sec:DR}

\begin{figure}
\begin{center}
\includegraphics[width=0.45\linewidth]{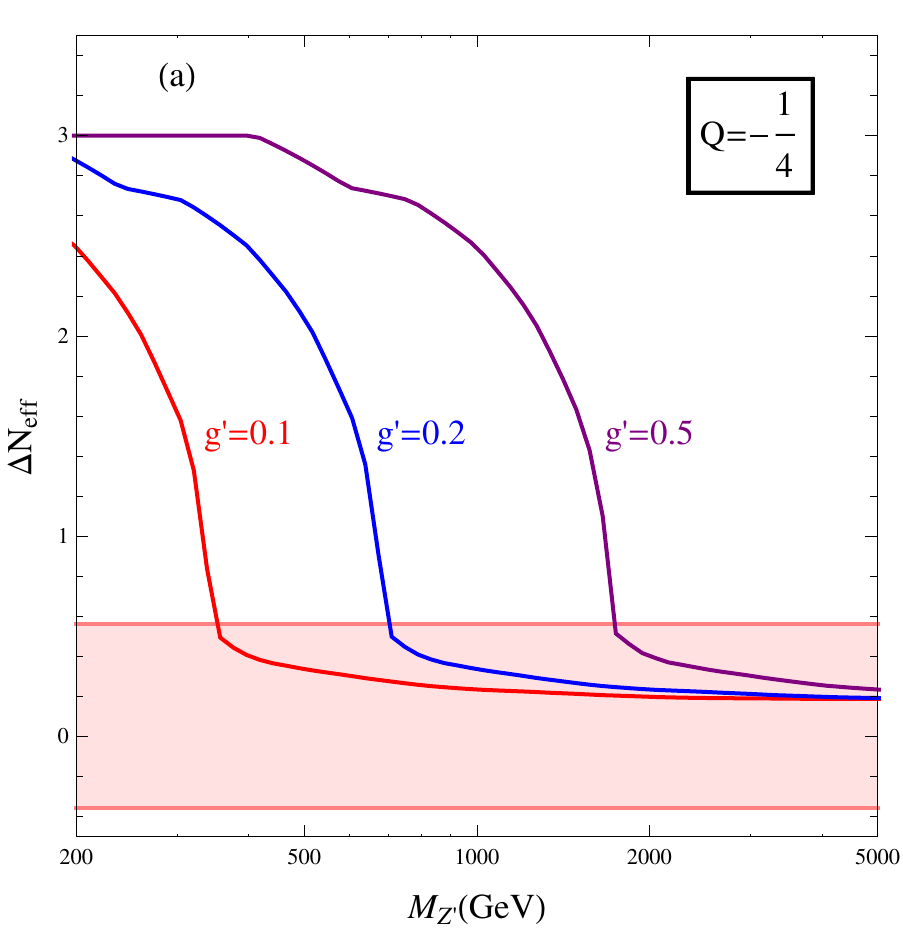}
\includegraphics[width=0.45\linewidth]{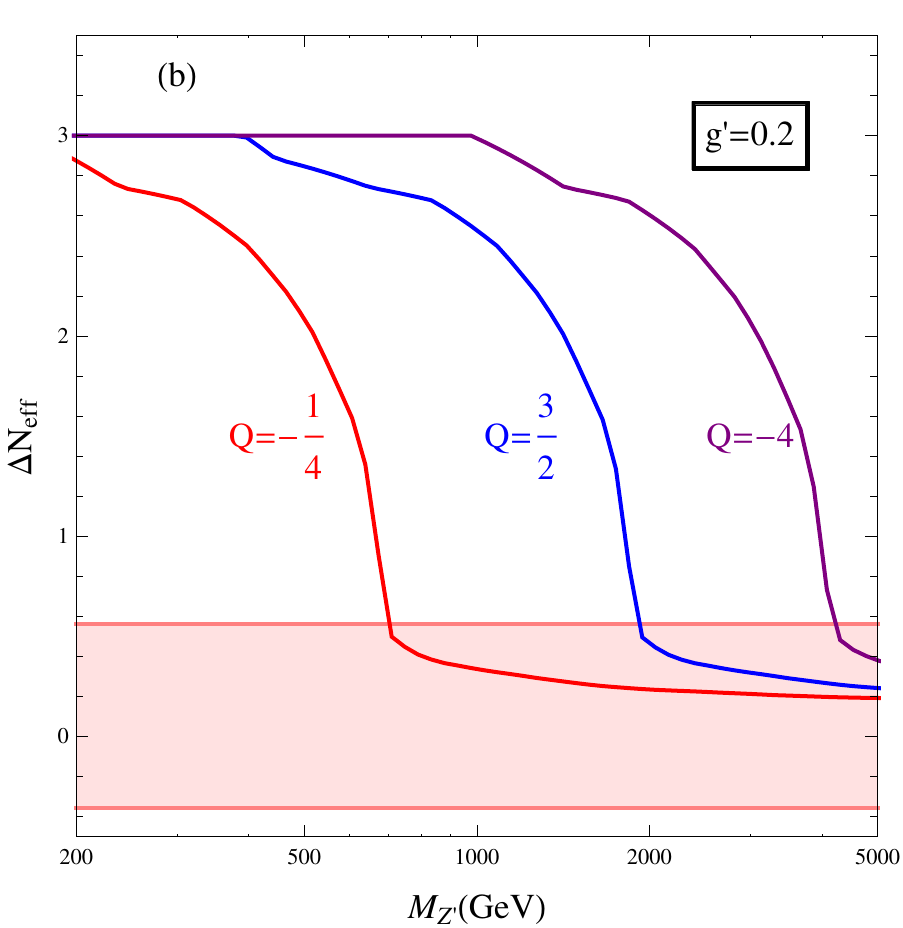}
\end{center}
\caption{$\Delta N_{\text{eff}}$ as a function of $M_{Z'}$. (a) $Q=-\frac{1}{4}$, $g'=0.1,0.2,0.5$. (b) $g'=0.2$, $Q=-\frac{1}{4}, -4, \frac{3}{2}$. The shaded pink region is allowed by current experiments \cite{Ade:2015xua}.
\label{Fig:DNv}}
\end{figure}

In this $B-L$ scotogenic Dirac model, the light right-handed neutrinos $\nu_R$ serve as a new form of dark radiation in the expansion of the universe, which will affect the effective number of relativistic degrees of freedom $N_\text{eff}$ for neutrinos. Here, we use the combined Planck TT+lowP+BAO data, $N_\text{eff}=3.15\pm0.23$\cite{Ade:2015xua}. One naively may worry that the existence of three generation of $\nu_R$ is in conflict with observed data, since the relativistic degrees of freedom for neutrinos are doubled. Actually, due to different gauge interactions, $\nu_R$ could decouple earlier than $\nu_L$, leading to a suppression of contribution to $N_\text{eff}$ from $\nu_R$.
The corresponding theoretical contribution of $\nu_R$ to $\Delta N_\text{eff}=N_\text{eff}-N_\text{eff}^\text{SM}$ is given by \cite{Anchordoqui:2011nh,Anchordoqui:2012qu}
\begin{equation}\label{Eq:Neff}
\Delta N_{\text{eff}}
= 3 \left(\frac{g(T^{\nu_L}_\text{dec})}{g(T^{\nu_R}_\text{dec})}\right)^{4/3},
\end{equation}
where $N_{\text{eff}}^\text{SM}=3.046$ comes from the contribution of SM left-handed neutrinos \cite{Mangano:2001iu}, and $g(T)$ is the effective number of degrees of freedom at temperature $T$ \cite{SolagurenBeascoa:2012cz}. The decoupling temperature of left-handed neutrinos $\nu_L$ is $T^{\nu_L}_\text{dec}\sim3~\MeV$, thus $g(T^{\nu_L}_\text{dec})=43/4$ corresponds to three $\nu_L$, $e^\pm$ and photon \cite{Kolb:1990vq}. In the gauged $U(1)_{B-L}$ scotogenic Dirac model, $\nu_R$ are in equilibrium with SM fermions via the new gauge boson $Z'$, therefore the interaction rate of $\nu_R$ is calculated as \cite{SolagurenBeascoa:2012cz}
\begin{eqnarray}\label{Eq:TvR}
\Gamma_{\nu_R}(T)&=&n_{\nu_R}(T)\sum_f \langle \sigma_f(\nu_R \bar{\nu}_R\to f\bar{f})v \rangle \\ \nonumber
&=& \sum_f \frac{g_{\nu_R}^2}{n_{\nu_R}}\int \frac{d^3 p }{(2\pi)^3} \frac{d^3 q }{(2\pi)^3}
f_{\nu_R}(p)f_{\nu_R}(q) \sigma_f(s) v,
\end{eqnarray}
where $s=2pq(1-\cos\theta)$, $v=(1-\cos\theta)$ with $\theta$ the relative angle of the colliding right-handed neutrinos. $f_{\nu_R}(p)$ is the Fermi-Dirac distribution function $f_{\nu_R}(k)=1/(e^{k/T}+1)$ and the number density of the right-handed neutrinos $n_{\nu_R}$ with spin number $g_{\nu_R}=2$ is given by
\begin{equation}
n_{\nu_R}(T)=g_{\nu_R}\int \frac{d^3k}{(2\pi)^3}f_{\nu_R}(k).
\end{equation}
In the limit $M^2_{Z'}\gg s$, the annihilation cross section in Eq.~\ref{Eq:TvR} is given by
\begin{equation}\label{Eq:sf}
\sigma_f(s)\simeq\frac{ N_C^f (Q_{BL}^f)^2 Q^2 s}{12\pi}\left(\frac{g'}{M_{Z'}}\right)^4.
\end{equation}
Including the contribution of three right-handed neutrinos, the Hubble parameter is now derived by
\begin{equation}
H(T)=\sqrt{\frac{4\pi^3G_N(g(T)+\frac{21}{4})}{45}}\,T^2.
\end{equation}

The right-handed neutrinos decouple when $\Gamma_{\nu_R}(T^{\nu_R}_\text{dec})=H(T^{\nu_R}_\text{dec})$. Solving this condition numerically, one can obtain $T^{\nu_R}_\text{dec}$, and then $\Delta N_\text{eff}$ via Eq.~\ref{Eq:Neff}. The results of $\Delta N_\text{eff}$ as a function of $M_{Z'}$ are exhibited in Fig.~\ref{Fig:DNv}. According to Eq.~\ref{Eq:sf}, for a larger $g'$, $|Q|$ or a smaller $M_{Z'}$, the corresponding annihilation cross section is larger, resulting in a smaller decoupling temperature $T^{\nu_R}_\text{dec}$ \cite{SolagurenBeascoa:2012cz}. Therefore, $g(T^{\nu_R}_\text{dec})$ is smaller and then $\Delta N_\text{eff}$ is larger. Such changes are clearly shown in Fig.~\ref{Fig:DNv}. We notice that for TeV-scale $Z'$, the minimum value of $\Delta N_\text{eff}$ is about 0.2, corresponding to a plateau around $g(T_\text{dec}^{\nu_R})\sim86$ when $3~\GeV\lesssim T_\text{dec}^{\nu_R}\lesssim10~\GeV$, and are hopefully within the reach of future Euclid experiment \cite{Amendola:2016saw}.

\begin{figure}
\begin{center}
\includegraphics[width=0.45\linewidth]{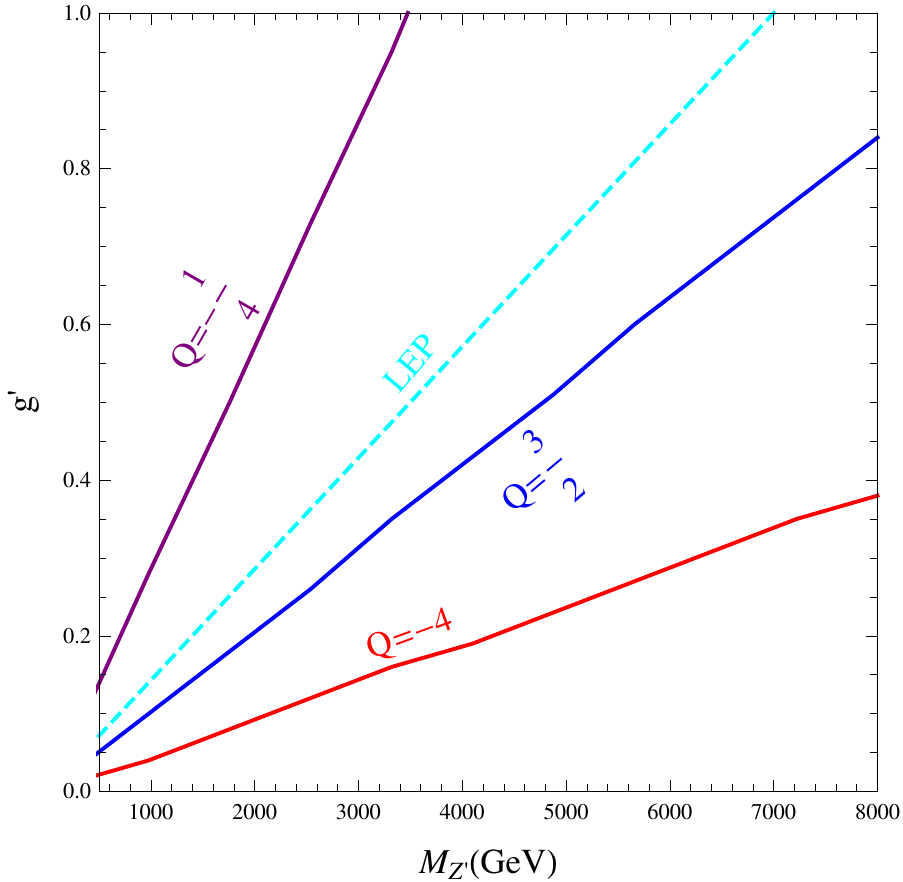}
\end{center}
\caption{$\Delta N_\text{eff}$ exclusion limit in the $g'-M_{Z'}$ for $Q=-\frac{1}{4},-4,\frac{3}{2}$. The dashed line corresponds to the LEP limit in Eq.~\ref{Eq:LEP}.
\label{Fig:gBL}}
\end{figure}

By requiring $\Delta N_\text{eff}=0.564$, one can further derive the exclusion limit in the $g'-M_{Z'}$ plane, which is shown in Fig.~\ref{Fig:gBL}. Depending on $|Q|$, the exclusion limits are approximately $M_{Z'}/g'\lesssim 3.5, 9.5, 21~\TeV$ for $Q=-\frac{1}{4},\frac{3}{2},-4$ respectively.
In contrast to LHC limits, a larger $|Q|$ will lead to a more stringent limit from $\Delta N_\text{eff}$. And for $Q=\frac{3}{2}$ and $-4$, the exclusion limits have already exceeded LEP limit.

\section{Dark Matter}\label{Sec:DM}
In this section, we investigate the phenomenology of $Z'$-portal DM, which is determined by relevant interactions presented in Eq.~\ref{Eq:LZp}. As stated in Sec.~\ref{Sec:MD}, we consider $H_1^0$ and $F_1$ as scalar and fermion DM candidate respectively. DM observables, such as the DM relic density, the DM-nucleon cross section and the thermally averaged annihilation cross section $\langle \sigma v \rangle$, are calculated with the help of {\tt micrOMEGAs} \cite{Belanger:2014vza} for a more precise results. In order to illustrate the dependence of DM observables on certain variables, we first carry out the results of some benchmark scenarios in this section. The combined analysis of different observables and a scan over corresponding parameter space will be performed in Sec.~\ref{Sec:CA}.

\subsection{Relic Density}\label{Sec:RD}

Here, we consider the conventional WIMP DM freeze-out scenario. As long as the temperature $T$ is high enough, the DM candidate is in thermal equilibrium with the primordial thermal bath via the new gauge interaction. And after the DM interaction rate is smaller than the expansion rate of the Universe, it is decoupled. The DM number density $n$ is then calculated by solving the following Boltzmann equation \cite{Bertone:2004pz}
\begin{equation}
\frac{d n}{dt} + 3 H n= -\langle \sigma v \rangle (n^2-n_\text{eq}^2),
\end{equation}
where $H$ is the Hubble parameter, and $n_\text{eq}$ is the thermal equilibrium value. In the following numerical results, we use {\tt micrOMEGAs} \cite{Belanger:2014vza} to solve the above Boltzmann equation and determine the DM relic density. Approximately, one can estimate the DM relic density by
\begin{equation}\label{Eq:RD}
\Omega h^2\approx \frac{3\times10^{-27} \text{cm}^3\text{s}^{-1}}{\langle\sigma v\rangle},
\end{equation}
which will be taken for qualitative illustration, since it is more intuitional. Moreover, the DM relic density measured by Planck is $\Omega h^2=0.1199\pm0.0027$ \cite{Ade:2015xua}.

In the $B-L$ scotogenic Dirac model, the DM candidate $H_1^0$ or $F_1$ can annihilate into $Z_2$-even fermion pairs via the exchange of $Z'$ in the $s$-channel. Neglecting masses of the final state fermions, the corresponding thermal averaged annihilation cross sections are given by \cite{Arcadi:2017kky}
\begin{equation}\label{Eq:svs1}
\langle\sigma v\rangle_{H_1^0 H_1^{0*}\to f\bar{f}}\simeq \frac{2 g'^4 N_C^f (V_f^2+A_f^2) Q^2_{H_1^0} M_{H_1^0}^2}{3\pi [(4M_{H_1^0}^2-M_{Z'}^2)^2+M_{Z'}^2\Gamma_{Z'}^2]} v^2,
\end{equation}
\begin{equation}\label{Eq:svf1}
\langle\sigma v\rangle_{F_1\bar{F}_1\to f\bar{f}}\simeq \frac{2g'^4 N_C^f(V_f^2+A_f^2) V_{F_1}^2 M_{F_1}^2}{\pi[(4M_{F_1}^2-M_{Z'}^2)^2+M_{Z'}^2\Gamma_{Z'}^2]},
\end{equation}
for scalar DM $H_1^0$ and fermion DM $F_1$ respectively. Here, $V_q=\frac{1}{3}$, $A_q=0$, $V_\ell=-1$, $A_\ell=0$, $V_\nu=-V_{F_1}=\frac{Q-1}{2}$, $A_\nu=-A_{F_1}=\frac{Q+1}{2}$, can be extracted from Eq.~\ref{Eq:LZp}. In the mean time, since the scalar DM $H_1^0$ is dominant by the $\chi$-component, we set $Q_{H_1^0}\simeq Q_\chi=Q-1$. Clearly, the annihilation cross sections for scalar DM is velocity suppressed. If DM is heavier than $Z'$, then DM pair can also annihilate into $Z'$ pair, leading to annihilation cross sections as \cite{Arcadi:2017kky}
\begin{eqnarray}\label{Eq:svs2}
\langle\sigma v\rangle_{H_1^0 H_1^{0*}\to Z'Z'} &\simeq& \frac{g'^4 Q_{H_1^0}^4 M_{H_1^0}^2}{16\pi M_{Z'}^4}\left(1-\frac{M_{Z'}^2}{M_{H_1^0}^2}\right)^{\frac{1}{2}}
\left(1-\frac{M_{Z'}^2}{2M_{H_1^0}^2}\right)^{-2} \\ \nonumber
&\times& \left[4\left(1-\frac{M_{Z'}^2}{M_{H_1^0}^2}\right)^2
-\left(4-4\frac{M_{Z'}^2}{M_{H_1^0}^2}-3\frac{M_{Z'}^4}{M_{H_1^0}^4}\right)
\left(1-\frac{M_{Z'}^2}{2M_{H_1^0}^2}\right)^2\right],
\end{eqnarray}
\begin{eqnarray}\label{Eq:svf2}
\langle\sigma v\rangle_{F_1\bar{F}_1\to Z'Z'} &\simeq& \frac{g'^4}{4\pi M_{F_1}^2 M_{Z'}^2}\left(1-\frac{M_{Z'}^2}{M_{F_1}^2}\right)^{\frac{3}{2}}
\left(1-\frac{M_{Z'}^2}{2M_{F_1}^2}\right)^{-2} \\ \nonumber
&\times&\left[(V_{F_1}^4+A_{F_1}^4)M_{Z'}^2+2V_{F_1}^2A_{F_1}^2(4M_{F_1}^2-3M_{Z'}^2)\right].
\end{eqnarray}
Therefore, free parameters involved in DM relic density are gauge coupling $g'$, $B-L$ charge $Q$, gauge boson mass $M_{Z'}$, and DM mass $M_{H_1^0,F_1}$, impacts of which are discussed in the following.

\begin{figure}
\begin{center}
\includegraphics[width=0.33\linewidth]{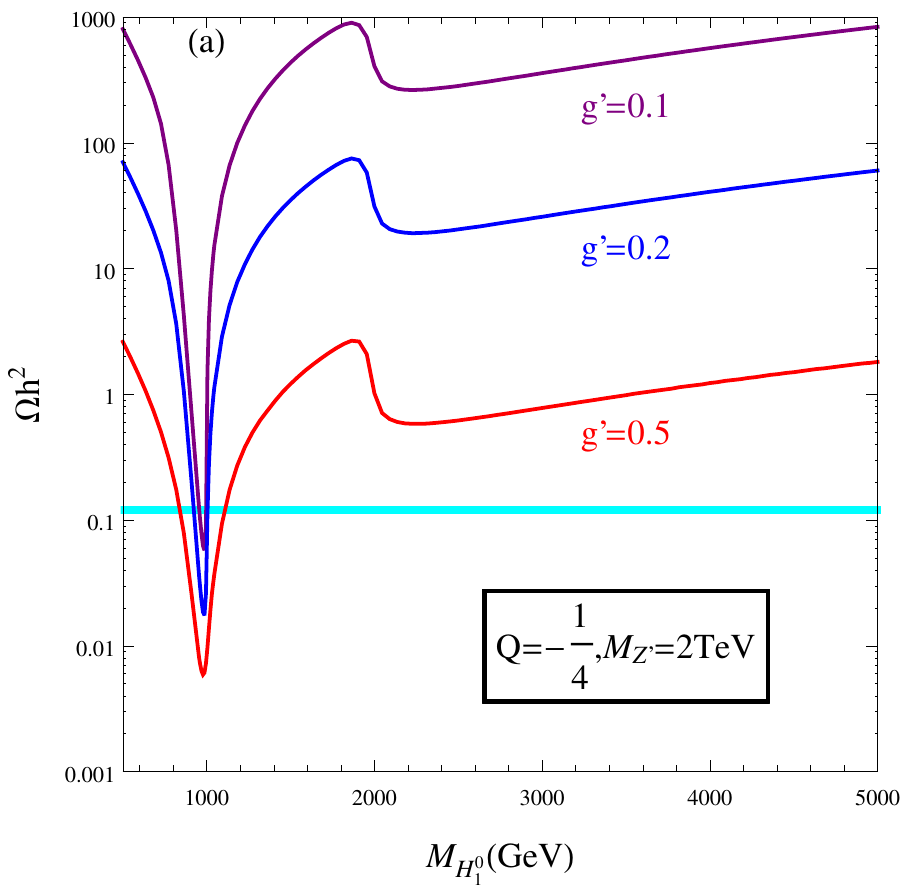}
\includegraphics[width=0.33\linewidth]{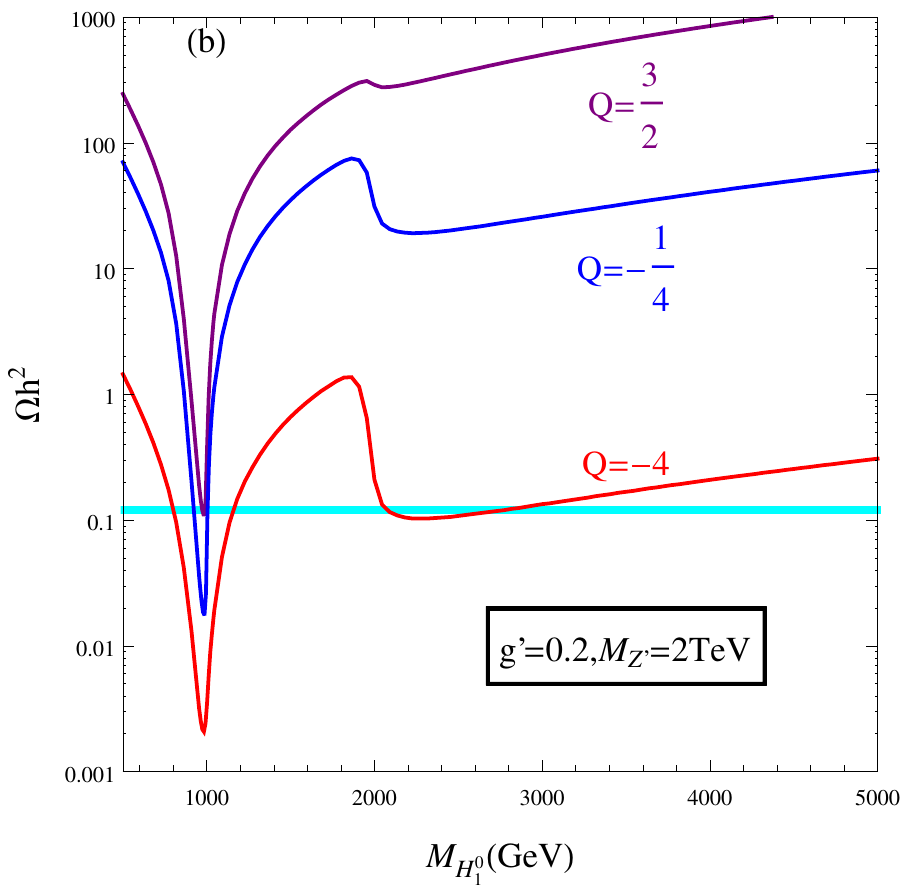}
\includegraphics[width=0.33\linewidth]{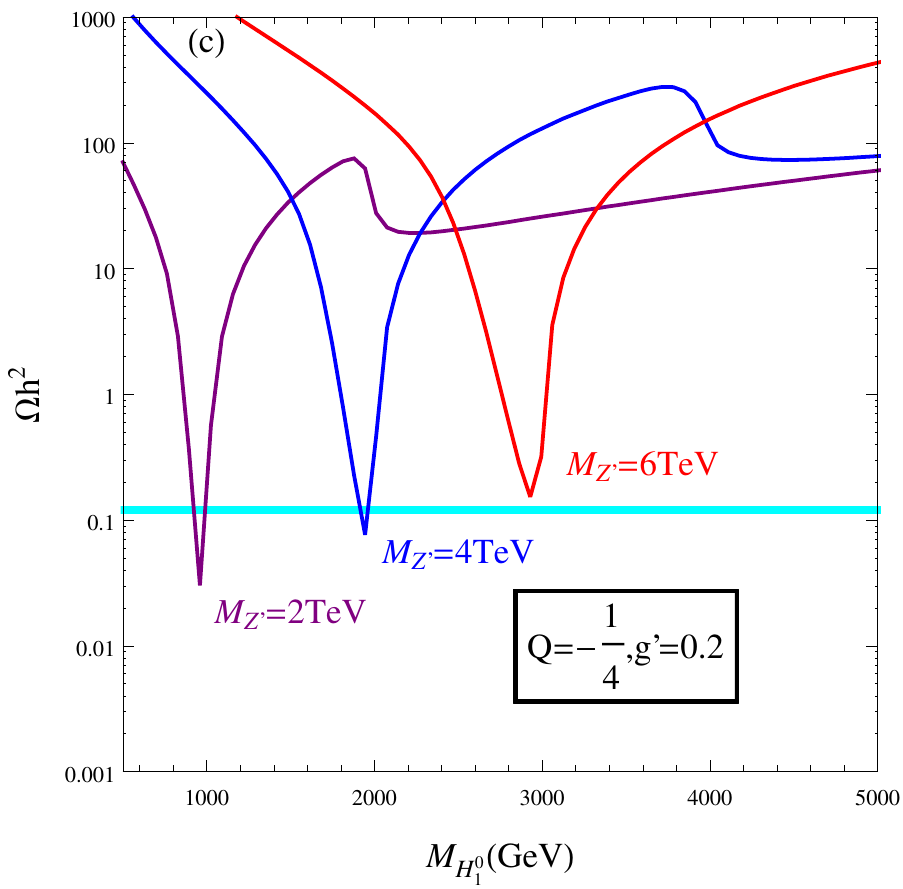}
\end{center}
\caption{Relic density as a function of scalar DM mass $M_{H_1^0}$. (a) $Q=-\frac{1}{4}$, $M_{Z'}=2~\TeV$, $g'=0.1,0.2,0.5$. (b) $g'=0.2$, $M_{Z'}=2~\TeV$, $Q=-\frac{1}{4},-4,\frac{3}{2}$.
(c) $Q=-\frac{1}{4}$, $g'=0.2$, $M_{Z'}=2,4,6~\TeV$. The cyan band corresponds to Planck measured result $\Omega h^2=0.1199\pm0.0027$ \cite{Ade:2015xua}.
\label{Fig:RDS}}
\end{figure}

First, we consider the relic density of scalar DM $H_1^0$, which is shown in Fig.~\ref{Fig:RDS}. Because of resonant production of $Z'$ in the $s$-channel, all the sub-figures exhibit sharp dips around $M_{H_1^0}\sim M_{Z'}/2$ in the relic density curves, which is a clear behaviour of Eq.~\ref{Eq:svs1}. And the resonance dips are broadened with increasing gauge $g'$, absolute value of $H_1^0$ charge $|Q_{H_1^0}|=|Q-1|$ and gauge boson mass $M_{Z'}$. In addition, around $M_{H_1^0}\sim M_{Z'}$, one also sees a decrease of relic density as increase of DM mass. Because, when $M_{H_1^0}>M_{Z'}$, the velocity independent annihilation process $H_1^0 H_1^{0*}\to Z'Z'$ (see Eq.~\ref{Eq:svs2}) is kinematically allowed and dominates over the velocity dependent process $H_1^0 H_1^{0*}\to Z'^* \to f\bar{f}$ (see Eq.~\ref{Eq:svs1}).

More specifically, in Fig.~\ref{Fig:RDS} (a), we fix $Q=-\frac{1}{4}$, $M_{Z'}=2~\TeV$, and vary $g'=0.1,0.2,0.5$. It is obvious that a larger $g'$ leads to a larger annihilation cross section, thus a smaller relic density. By requiring $\Omega h^2\simeq0.12$ to satisfy observed value, one actually finds two viable DM mass $M_{H_1^0}$, one below $M_{Z'}/2$ and the other above $M_{Z'}/2$. In Fig.~\ref{Fig:RDS} (b), $g'=0.2$, $M_{Z'}=2~\TeV$ are fixed and $Q=-\frac{1}{4},-4,\frac{3}{2}$ is varied. From Eq.~\ref{Eq:svs1} and Eq.~\ref{Eq:svs2}, one is aware that the annihilation cross sections are positively correlated with $|Q_{H_1^0}|=|Q-1|$. So $Q=\frac{3}{2}$ is equivalent to $Q_{H_1^0}=\frac{1}{2}$, corresponding to smallest $|Q_{H_1^0}|$ and largest relic density. Notably, for $Q=-4$, i.e., $Q_{H_1^0}=-5$, $g'Q_{H_1^0}=-1$, the cross section of annihilation process $H_1^0 H_1^{0*}\to Z'Z'$ is sufficient large to reach $\Omega h^2\simeq0.12$. Fig.~\ref{Fig:RDS} (c) depicts the impact of gauge boson mass $M_{Z'}$ when fixing $Q=-\frac{1}{4}$ and $g'=0.2$. Although the resonance dips exist for all kinds of $M_{Z'}$, the minimum value of relic density increases as $M_{Z'}$ increases. And eventually for $M_{Z'}=6~\TeV$, the minimum value of relic density is larger than the observed value $\sim0.12$.

\begin{figure}
\begin{center}
\includegraphics[width=0.33\linewidth]{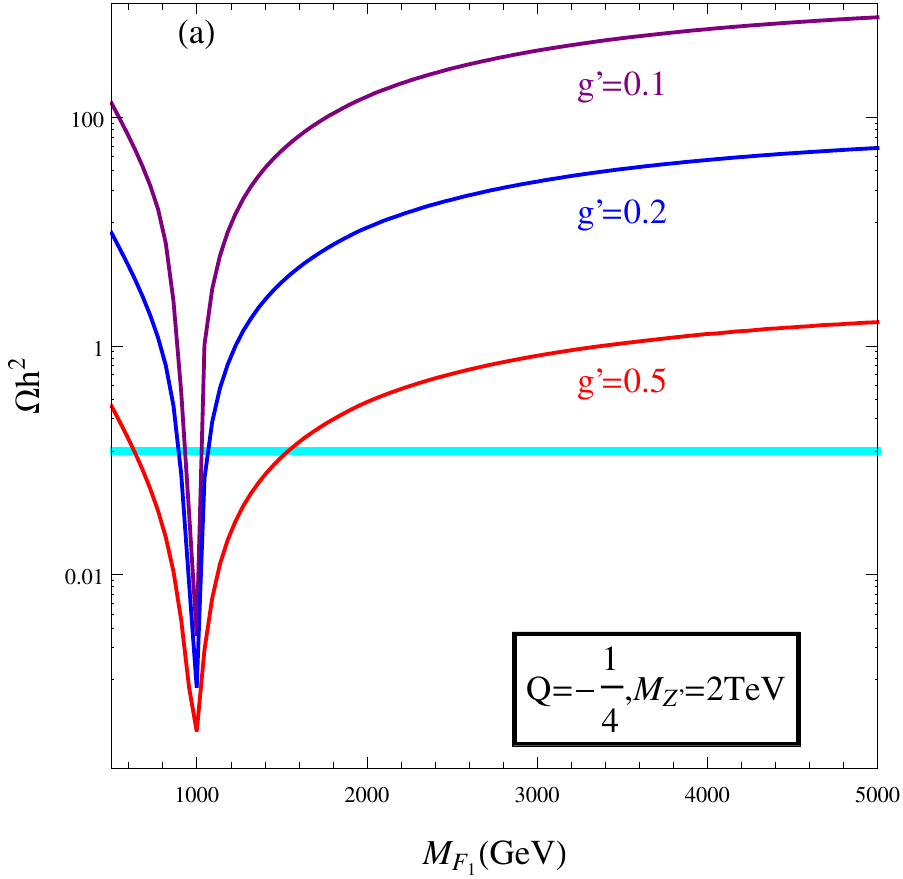}
\includegraphics[width=0.33\linewidth]{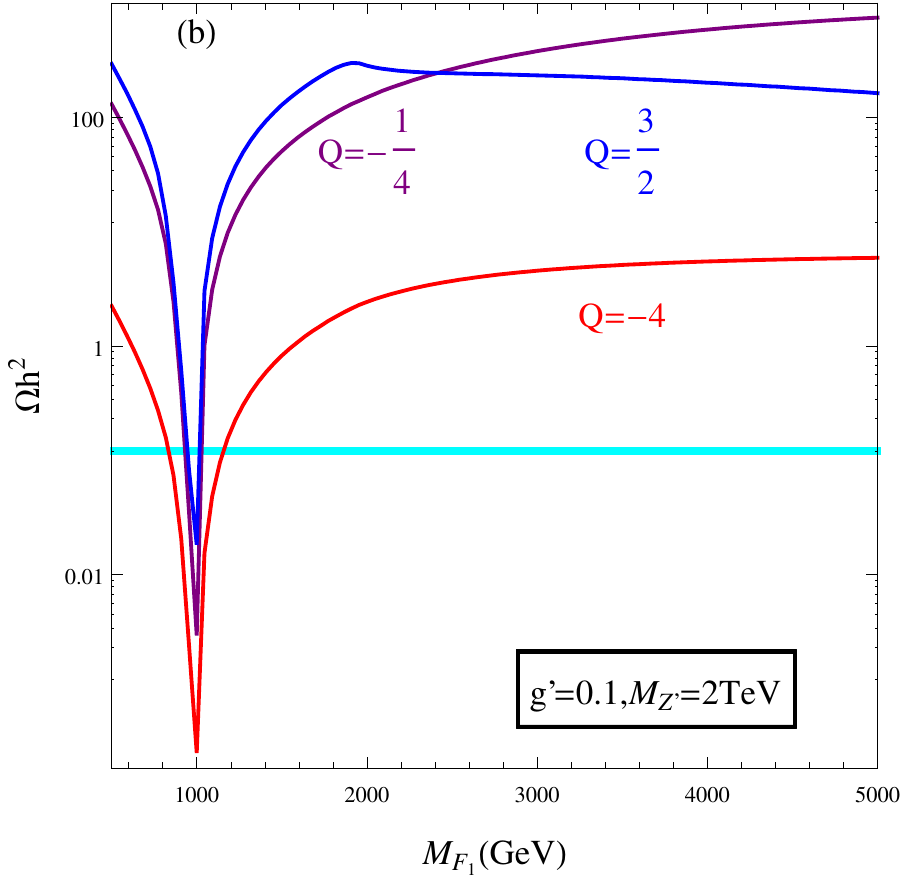}
\includegraphics[width=0.33\linewidth]{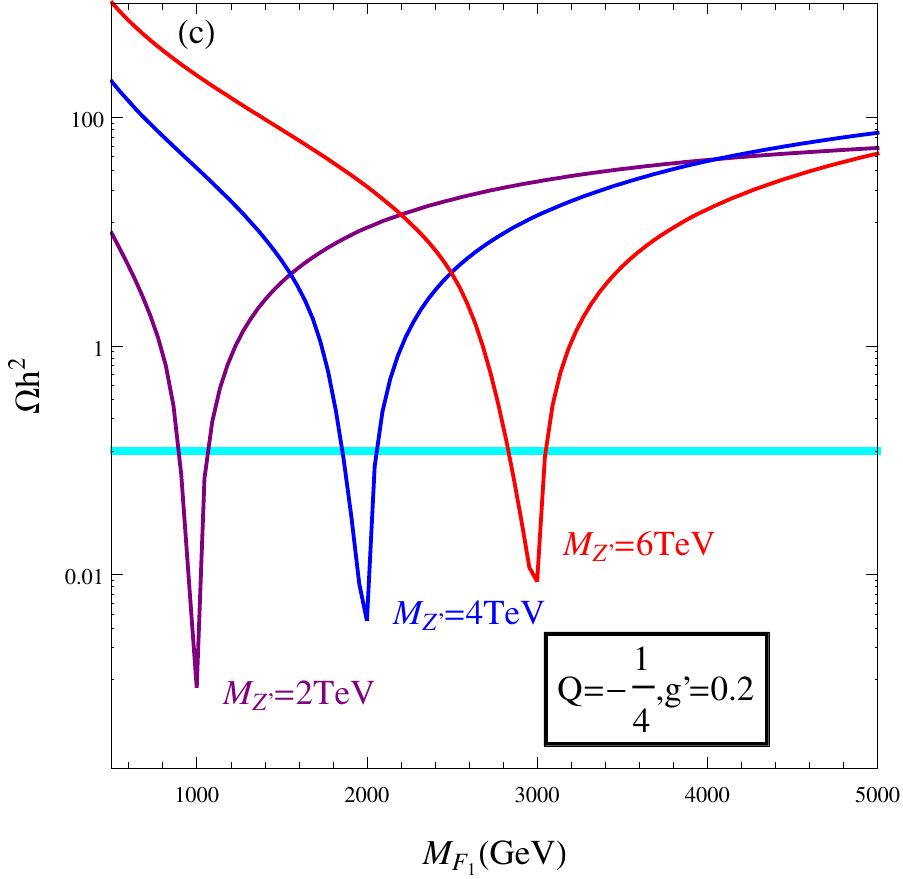}
\end{center}
\caption{Relic density as a function of fermion DM mass $M_{F_1}$. (a) $Q=-\frac{1}{4}$, $M_{Z'}=2~\TeV$, $g'=0.1,0.2,0.5$. (b) $g'=0.1$, $M_{Z'}=2~\TeV$, $Q=-\frac{1}{4},-4,\frac{3}{2}$.
(c) $Q=-\frac{1}{4}$, $g'=0.2$, $M_{Z'}=2,4,6~\TeV$. The cyan band corresponds to Planck measured result $\Omega h^2=0.1199\pm0.0027$ \cite{Ade:2015xua}.
\label{Fig:RDF}}
\end{figure}

Then, we move on to fermion DM. The results are shown in Fig.~\ref{Fig:RDF}. Similar to scalar DM, resonance dips appear around $M_{F_1}\sim M_{Z'}/2$. But comparing with scalar DM, the relic density of fermion DM is easier to reach $\Omega h^2\sim0.12$ with same values of free parameters, because the $s$-channel process $F_1\bar{F}_1\to Z'^*\to f\bar{f}$ is not velocity suppressed (see Eq.~\ref{Eq:svs1} and Eq.~\ref{Eq:svf1}). In the limit of $M_{F_1}^2\gg M_{Z'}^2$, one approximately has
\begin{equation}
\langle\sigma v\rangle_{F_1\bar{F}_1\to f\bar{f}}\sim \frac{g'^4 N_C^f(V_f^2+A_f^2) V_{F_1}^2}
{8\pi M_{F_1}^2} \ll
\langle\sigma v\rangle_{F_1\bar{F}_1\to Z'Z'} \sim \frac{2g'^4 V_{F_1}^2 A_{F_1}^2}{\pi M_{Z'}^2}.
\end{equation}
So with increasing $M_{F_1}$, the process $F_1\bar{F}_1\to Z'Z'$ gradually becomes more important. Besides, the annihilation cross section, thus the resulting relic density, tends to a constant when $M_{F_1}$ increases.

The impacts of gauge coupling $g'$, $\nu_R$'s $B-L$ charge $Q$ and gauge boson mass $M_{Z'}$ are shown in Fig.~\ref{Fig:RDF} (a), (b) and (c) respectively. Again, effects of changing $g'$ and $M_{Z'}$ are similar as scalar DM, but not $Q$. When $M_{F_1}<M_{Z'}$, Eq.~\ref{Eq:svf1} indicates that the vector coupling $V_{F_1}$ is the dominant contribution. So $Q=-\frac{1}{4}$  with $|V_{F_1}|=\frac{5}{8}$ corresponds to larger cross section and smaller relic density than $Q=\frac{3}{2}$ with $|V_{F_1}|=\frac{1}{4}$. But when $M_{F_1}>M_{Z'}$, one has to take into account the contribution of axial-vector coupling $A_{F_1}$ according to Eq.~\ref{Eq:svf2}. In this way, $Q=\frac{3}{2}$ leads to a larger annihilation cross section and a smaller relic density than $Q=-\frac{1}{4}$.

\subsection{Direct Detection}\label{Sec:DD}

Through $t$-channel exchange of $Z'$, the new gauge interactions in Eq.~\ref{Eq:LZp} would induce spin-independent scattering of DM on nucleon. Currently, various direct detection experiments \cite{Akerib:2016vxi,Aprile:2017iyp,Cui:2017nnn} have already set stringent bounds on the spin-independent scattering cross section. For $M_{\text{DM}}\lesssim100~\GeV$, the most stringent limit is provided by XENON1T2017~\cite{Aprile:2017iyp}, meanwhile PandaX2017~\cite{Cui:2017nnn} sets the most stringent limit for $M_{\text{DM}}\gtrsim100~\GeV$. Moreover, we also incorporate the projected limit by XENON1T(2t$\cdot$y)~\cite{Aprile:2015uzo} to illustrate future direct detection limit. Typically, for $M_\text{DM}\sim1~\TeV$, PandaX2017~\cite{Cui:2017nnn}  requires $\sigma^\text{SI}\lesssim10^{-9}$ pb, and XENON1T(2t$\cdot$y) will push this limit down to about $10^{-10}$ pb~\cite{Aprile:2015uzo}.

The spin-independent scattering cross section is given by \cite{DEramo:2016gos}
\begin{equation}\label{Eq:SI}
\sigma^\text{SI}\approx\frac{\mu_N^2}{\pi}\frac{g'^4}{M_{Z'}^4} Q^2_\text{DM}
=1.24\times10^{-4}\left(\frac{\mu_N}{1~\GeV}\right)^2\left(\frac{1~\TeV}{M_{Z'}}\right)^4
g'^4 Q_\text{DM}^2~(\pb),
\end{equation}
where $\mu_N=M_N M_\text{DM}/(M_N+M_\text{DM})$ is the reduced mass, $M_N=(M_n+M_p)/2=939~\MeV$ is the averaged nucleon mass, $Q_\text{DM}=Q_{H_1^0}=Q-1$ for scalar DM $H_1^0$, and $Q_\text{DM}=V_{F_1}=(1-Q)/2$ for fermion DM $F_1$. Provided $g'\sim0.1$, $M_{Z'}\sim1$, $Q_\text{DM}\sim1$ and $M_\text{DM}\sim1~\TeV$, then $\sigma^\text{SI}\sim10^{-8}~\pb$ is obtained, which is an order of magnitude larger than PandaX2017 limits ~\cite{Cui:2017nnn}. Because the spin-independent cross section is proportional to $Q_\text{DM}^2 g'^4/M_{Z'}^4$, the results of other free parameter sets can be easily derived by a simple rescaling. So in order to escape the direct detection limits, one naively expects that smaller $Q_\text{DM}$, smaller $g'$ and larger $M_{Z'}$ is required. But one also has to keep in mind that such choice of free parameters usually leads to a larger relic density as discussed in Sec.~\ref{Sec:RD}. Hence, for a more comprehensive analysis, we further require the free parameter set yielding correct relic density.

\begin{figure}
\begin{center}
\includegraphics[width=0.45\linewidth]{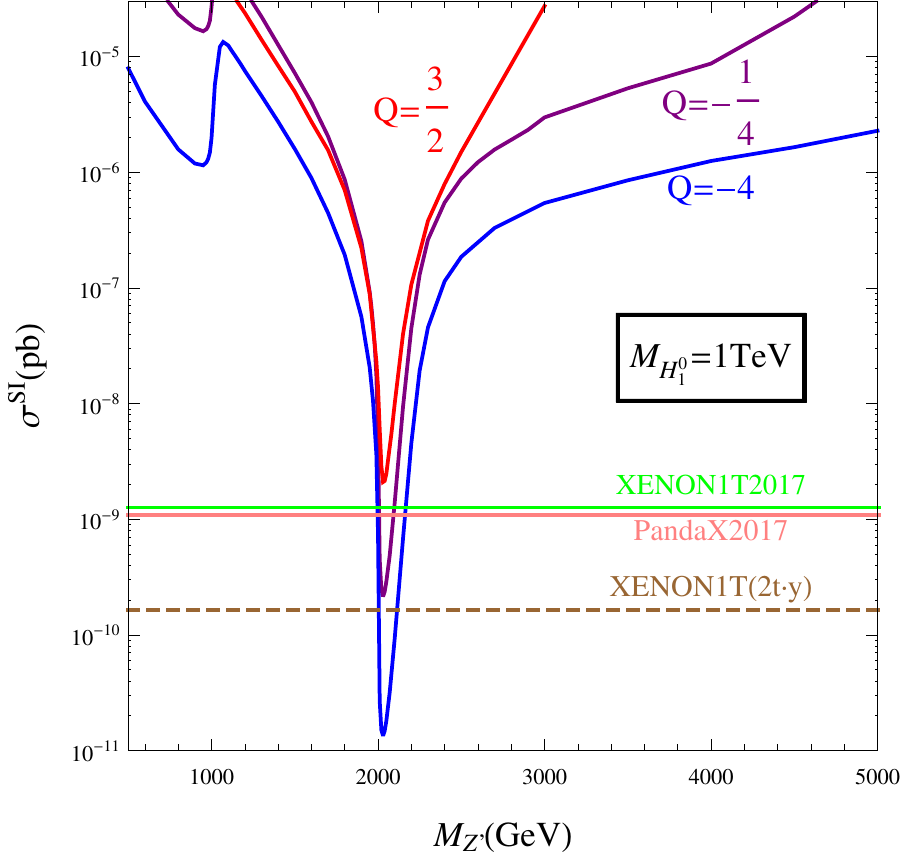}
\includegraphics[width=0.445\linewidth]{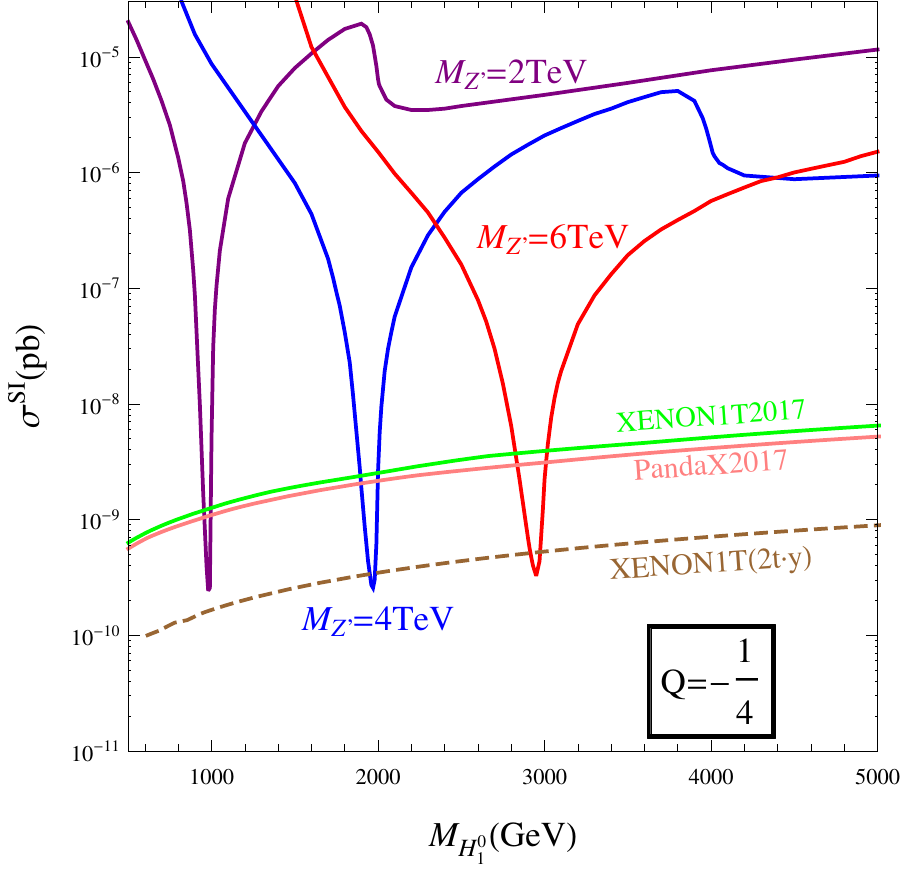}
\end{center}
\caption{Spin-independent cross section of scalar DM $H_1^0$ scattering on nucleon. Left: as a function of $M_{Z'}$ for $M_{H_1^0}=1~\TeV$, $Q=-\frac{1}{4},-4,\frac{3}{2}$. Right: as a function of $M_{H_1^0}$ for $M_{Z'}=2,4,6~\TeV$, $Q=-\frac{1}{4}$. Here, $g'$ is determined by  yielding correct relic density $\Omega h^2\simeq0.12$. Experimental bounds come from XENON1T2017(green)~\cite{Aprile:2017iyp}, PandaX2017(pink)~\cite{Cui:2017nnn}, and XENON1T(2t$\cdot$y)(brown)~\cite{Aprile:2015uzo}.
\label{Fig:DDS}}
\end{figure}

Again, we consider scalar DM first. In Fig.~\ref{Fig:DDS}, we display the predicted spin-independent cross section $\sigma^\text{SI}$. Left panel of Fig.~\ref{Fig:DDS} shows $\sigma^\text{SI}$ as a function of $M_{Z'}$, while we fix $M_{H_1^0}=1~\TeV$ and vary $Q=-\frac{1}{4},-4,\frac{3}{2}$. Similar as relic density, resonance dips also exist when $M_{H_1^0}\sim M_{Z'}/2$. Out of this resonance region, current direct detection experiments XENON1T2017~\cite{Aprile:2017iyp} and PandaX2017~\cite{Cui:2017nnn} have already excluded the corresponding parameter space, which indicates that the annihilation channel $H_1^0 H_1^{0*}\to Z'Z'$ is ruled out. Considering the impact of $B-L$ charge $Q_{H_1^0}=Q-1$,
one can see that a smaller $|Q_{H_1^0}|$ actually leads to larger DM-nucleon cross section when imposing relic density conditions. For $Q=\frac{3}{2}(|Q_{H_1^0}|=\frac{1}{2})$, it has been fully excluded by PandaX2017~\cite{Cui:2017nnn}. For $Q=-\frac{1}{4}(|Q_{H_1^0}|=\frac{5}{4})$, although the resonance region can escape current PandaX2017 limit~\cite{Cui:2017nnn}, it is within the reach of future XENON1T(2$\cdot$y)~\cite{Aprile:2015uzo}. For $Q=-4(|Q_{H_1^0}|=5)$, it can even avoid future XENON1T(2$\cdot$y) limit~\cite{Aprile:2015uzo} in the resonance region. Right panel of Fig.~\ref{Fig:DDS} depicts $\sigma^\text{SI}$ as a function of $M_{H_1^0}$, while we fix $Q=-\frac{1}{4}$ and vary $M_{Z'}=2,4,6~\TeV$. Basically, the larger $M_{Z'}$ is, the easier it will be to escape direct detection bounds, due to the bounds is less stringent with increasing DM mass. For $M_{Z'}=4,6~\TeV$, except for the region quite close to the resonance region $M_{H_1^0}\simeq M_{Z'}/2$, XENON1T(2$\cdot$y)~\cite{Aprile:2015uzo} could probe most of the parameter space.

\begin{figure}
\begin{center}
\includegraphics[width=0.45\linewidth]{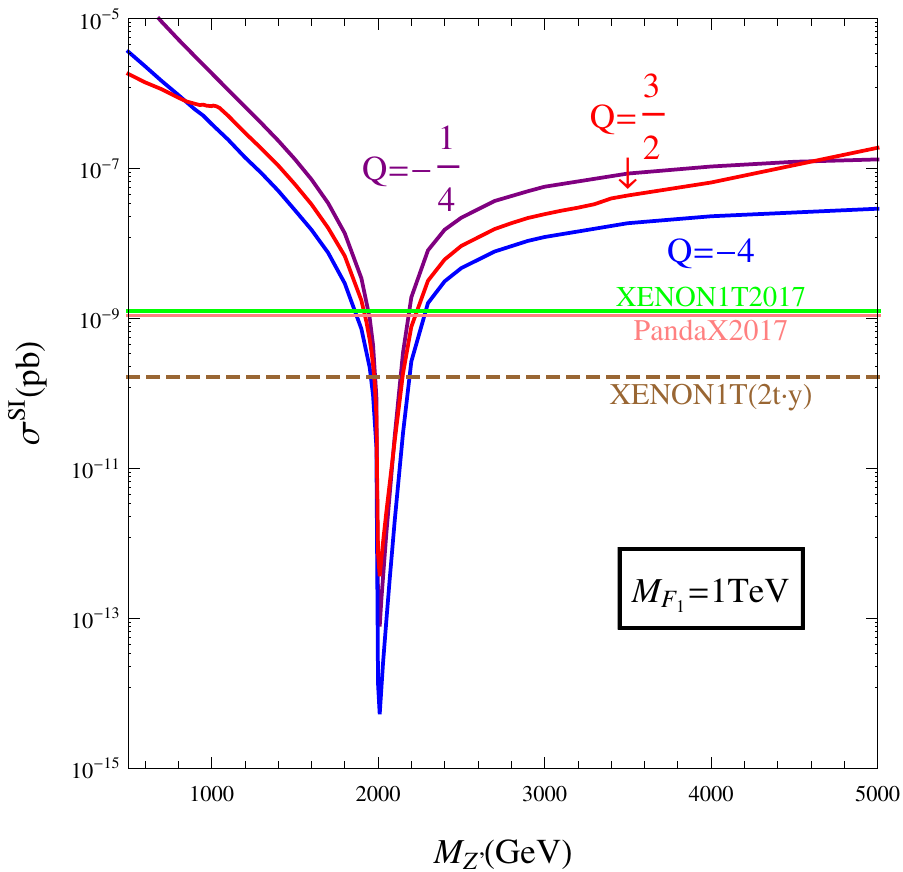}
\includegraphics[width=0.45\linewidth]{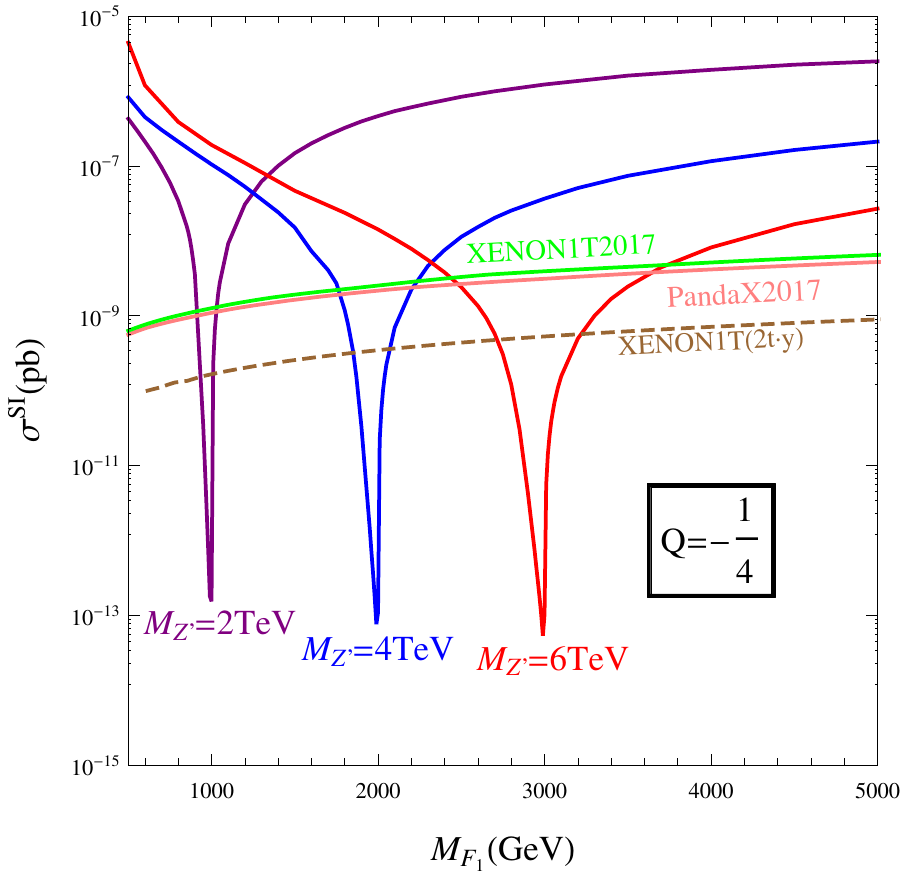}
\end{center}
\caption{Same as Fig.~\ref{Fig:DDS}, but for fermion DM $F_1$.
\label{Fig:DDF}}
\end{figure}

Then, we come to fermion DM $F_1$. The results are shown in Fig.~\ref{Fig:DDF}. Similar as scalar DM, only the resonance region can escape direct detection limits. But in this region, the resulting $\sigma^\text{SI}$ is much smaller, so that both varying $Q=-\frac{1}{4},-4,\frac{3}{2}$ and varying $M_{Z'}=2,4,6~\TeV$ could easily avoid current and even future limits. Since only the vector coupling $V_{F_1}=(1-Q)/2$ is involved in relic density (see Eq.~\ref{Eq:svs1}) and spin-independent DM-nucleon scattering (see Eq.~\ref{Eq:SI}),  $Q=\frac{3}{2}(|V_{F_1}|=\frac{1}{4})$ corresponds to the smallest $B-L$ charge. Hence, the largest $g'$ is required to reproduce correct relic density, leading to largest scattering cross section.

In summary, only the resonance region $M_\text{DM}\sim M_{Z'}/2$ can satisfy direct detection limits by requiring the free parameter set yielding correct relic density for both scalar and fermion DM. Moreover, in the resonance region, we find that a larger effective absolute value of $B-L$ charge, i.e., $|Q_{H_1^0}|=|Q-1|$ for scalar DM and $|Q_{F_1}|=|V_{F_1}|=|(1-Q)/2|$ for fermion DM, actually leads to a smaller spin-independent cross section. And comparing with scalar DM, fermion DM is easier to escape direct detection limits.

\subsection{Indirect Detection}\label{Sec:ID}

According to the results of direct detection in Sec.~\ref{Sec:DD}, the so-called secluded DM scenario $M_\text{DM}>M_{Z'}$ \cite{Pospelov:2007mp,Profumo:2017obk} has already been excluded. So in the $B-L$ scotogenic Dirac model, DM pair can only annihilate into SM fermion pair through the gauge boson $Z'$ in the $s$-channel nowadays. Then, this leads to high energy gamma-rays, which is detectable at
Fermi-LAT~\cite{Ackermann:2015zua}, H.E.S.S.~\cite{Abdallah:2016ygi,Abramowski:2011hc}, and the forthcoming CTA~\cite{Consortium:2010bc,Lefranc:2015pza}. Provided DM 100\% annihilating into $b\bar{b}$ or $\tau^+\tau^-$ with an annihilation cross section of $3\times10^{-26}~\text{cm}^3\text{s}^{-1}$, Fermi-LAT has excluded DM mass below $100~\GeV$~\cite{Ackermann:2015zua}. As for TeV-scale DM, H.E.S.S. has excluded $500~\GeV\lesssim M_\text{DM}\lesssim 2~\TeV$ when DM solely annihilates into $\tau^+\tau^-$ final states also assuming an annihilation cross section of $3\times10^{-26}~\text{cm}^3\text{s}^{-1}$~\cite{Abdallah:2016ygi,Abramowski:2011hc}. To illustrate prospect of indirect detection, we also take into account the upcoming CTA experiment~\cite{Consortium:2010bc} and considering the most optimistic limits in Ref.~\cite{Lefranc:2015pza}, which in principle could probe $M_\text{DM}\lesssim10~\TeV$ for a 100\% branching ratio into $\tau^+\tau^-$ and an annihilation cross section of $3\times10^{-26}~\text{cm}^3\text{s}^{-1}$. Since for TeV-scale DM, the $\tau^+\tau^-$ channel provides the most stringent limit, we will take this channel to illustrate for simplicity. A more appropriate pathway is taking into all annihilation channels \cite{Alves:2015mua,Patra:2016ofq}.

The differential photon flux in any angular direction can be obtained by
\begin{equation}
\frac{d \Phi_\gamma(\Delta \Omega)}{d E}(E_\gamma)= \frac{1}{4\pi} \frac{\langle\sigma v\rangle}{2 M_\text{DM}^2}\sum_f \text{BR}_f \frac{d N_\gamma^f}{d E_\gamma}\cdot J_{ann},
\end{equation}
where BR$_f$ is the branching ratio of $Z'$ into fermion pair $f$, $d N_\gamma^f/d E_\gamma$ is the differential $\gamma$-spectrum from fermion pair $f$, $J_{ann}$ is the astrophysical $J$-factor. And the annihilation cross sections are presented in Eq.~\ref{Eq:svs1} for scalar DM and Eq.~\ref{Eq:svf1} for fermion DM, respectively. Because the annihilation cross section today is velocity suppressed for scalar DM, the corresponding indirect detection signatures are usually neglected~\cite{Rodejohann:2015lca}. As for fermion DM, the annihilation cross section is not suppressed, hence indirect detection signatures are in principle  promising. Furthermore, since only the resonance region $M_\text{DM}\sim M_{Z'}/2$ is viable under constraints from relic density and direct detection, the naive estimation of DM annihilation cross section with Eq.~\ref{Eq:RD}, i.e., $\langle \sigma v \rangle\simeq3\times10^{-26}~\text{cm}^3\text{s}^{-1}$, is invalid \cite{Griest:1990kh,Gondolo:1990dk}. In this case, the annihilation cross section could be enhanced via the Breit-Wigner mechanisms~\cite{Ibe:2008ye,Guo:2009aj}, so as a consequence the indirect detection limits are strengthened.

\begin{figure}
\begin{center}
\includegraphics[width=0.45\linewidth]{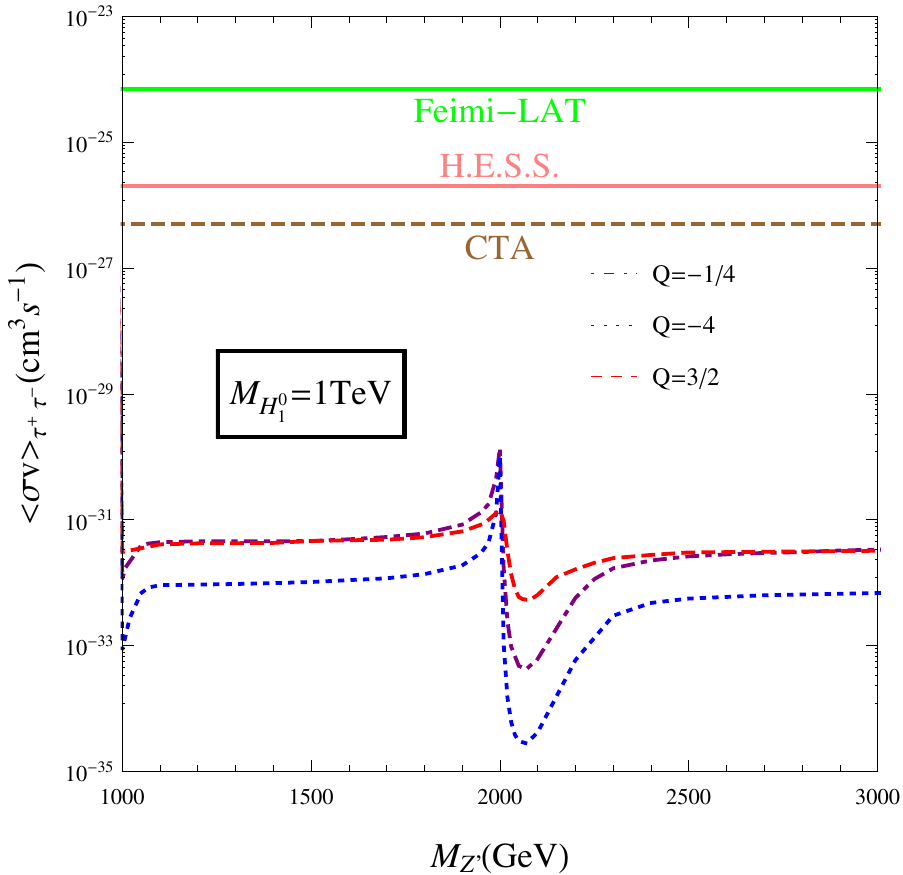}
\includegraphics[width=0.445\linewidth]{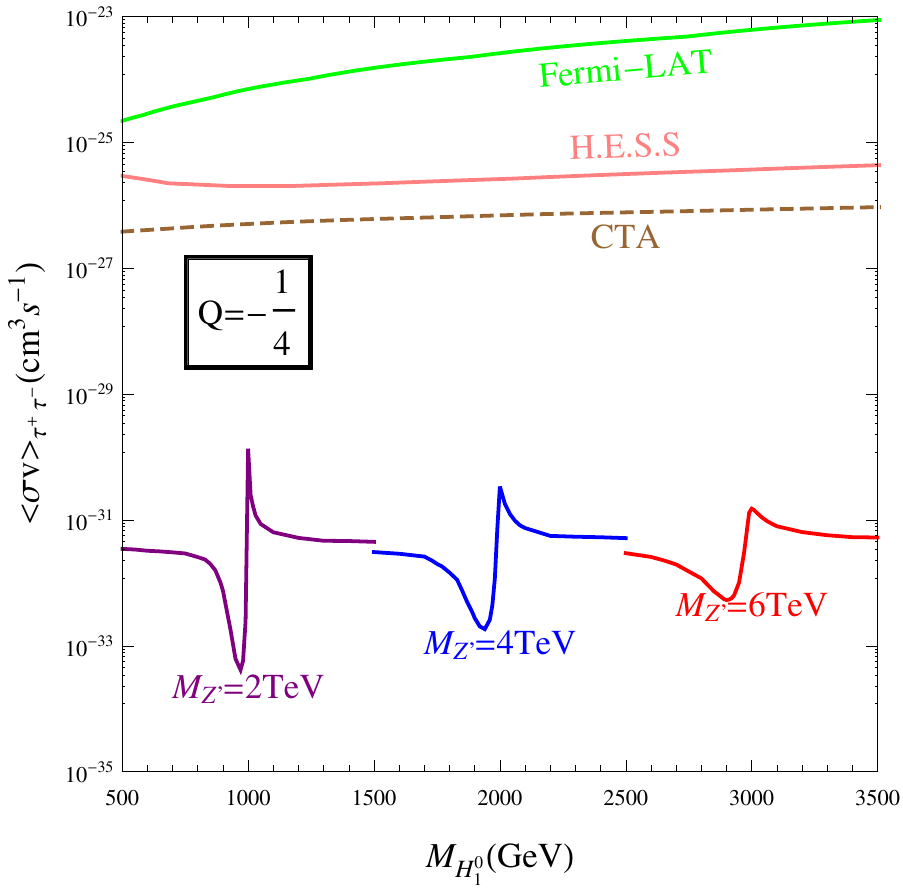}
\end{center}
\caption{Annihilation cross section for scalar DM $H_1^0$ into $\tau^+\tau^-$ final state. Left: as a function of $M_{Z'}$ for $M_{H_1^0}=1~\TeV$, $Q=-\frac{1}{4},-4,\frac{3}{2}$. Right: as a function of $M_{H_1^0}$ for $M_{Z'}=2,4,6~\TeV$, $Q=-\frac{1}{4}$. Here, $g'$ is determined by  yielding correct relic density $\Omega h^2\simeq0.12$. Experimental bounds are derived from Fermi-LAT(green)~\cite{Ackermann:2015zua}, H.E.S.S.(pink)~\cite{Abdallah:2016ygi}, and CTA(brown)~\cite{Lefranc:2015pza}.
\label{Fig:IDS}}
\end{figure}

We start with scalar DM as well. The annihilation cross sections for the $\tau^+\tau^-$ channel are shown in Fig.~\ref{Fig:IDS}. It is obvious that even with Breit-Wigner enhancement, the annihilation cross sections are still several orders of magnitudes lower than future CTA limits. In left panel of Fig.~\ref{Fig:IDS}, we show the impact of $B-L$ charge $Q$. Off the resonance region, one has a similar cross section for $Q=-\frac{1}{4}$ and $Q=\frac{3}{2}$, while the cross section is much smaller for $Q=-4$ due to sizable suppression of $\tau^+\tau^-$ branching ratio. When the gauge boson $Z'$ is almost on-shell, the cross section could reach about $2\times10^{-30}\text{cm}^3\text{s}^{-1}$, $1\times10^{-30}\text{cm}^3\text{s}^{-1}$ and $1\times10^{-31}\text{cm}^3\text{s}^{-1}$ for $Q=-\frac{1}{4}$, $Q=-4$ and $Q=\frac{3}{2}$ respectively. In right panel of Fig.~\ref{Fig:IDS}, we show $\langle\sigma v\rangle_{\tau^+\tau^-}$ as a function of $M_{H_1^0}$ by fixing $Q=-\frac{1}{4}$ and varying $M_{Z'}=2,4,6~\TeV$.  Here, we also impose $|M_{H_1^0}-M_{Z'}/2|\lesssim500~\GeV$ to show corresponding resonance region. Although the cross sections are similar when $M_{H_1^0}$ far away from $M_{Z'}$, we find that as $M_{Z'}$ increases, the maximum(minimum) value decreases(increases), and the resonance width becomes larger.

\begin{figure}
\begin{center}
\includegraphics[width=0.45\linewidth]{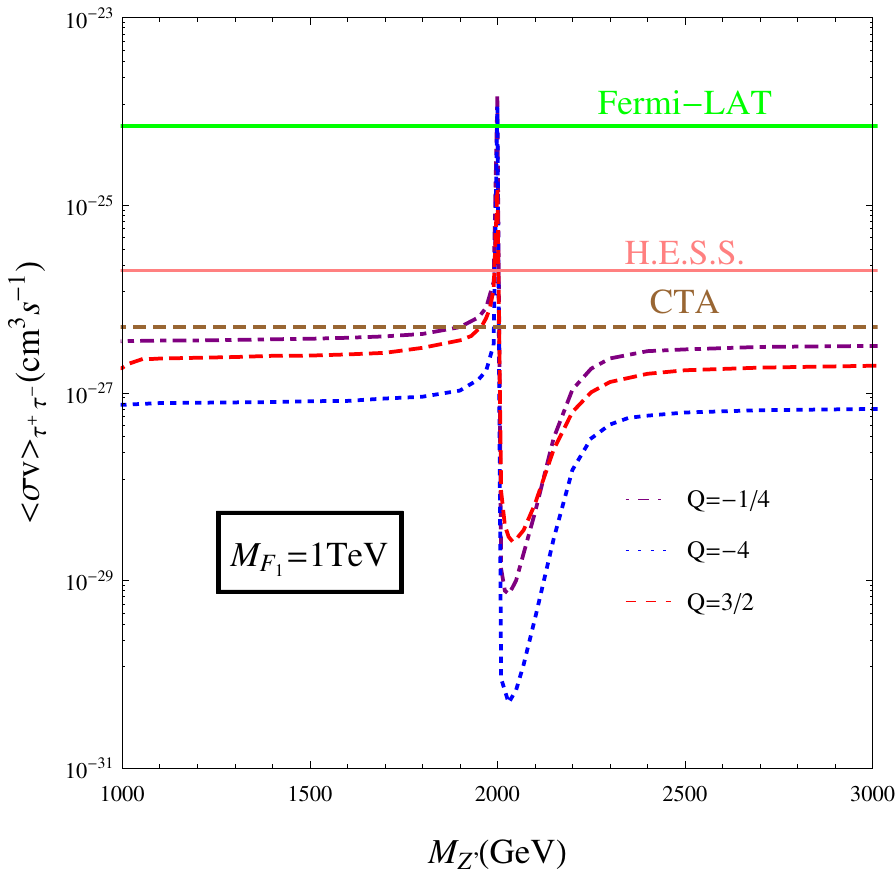}
\includegraphics[width=0.45\linewidth]{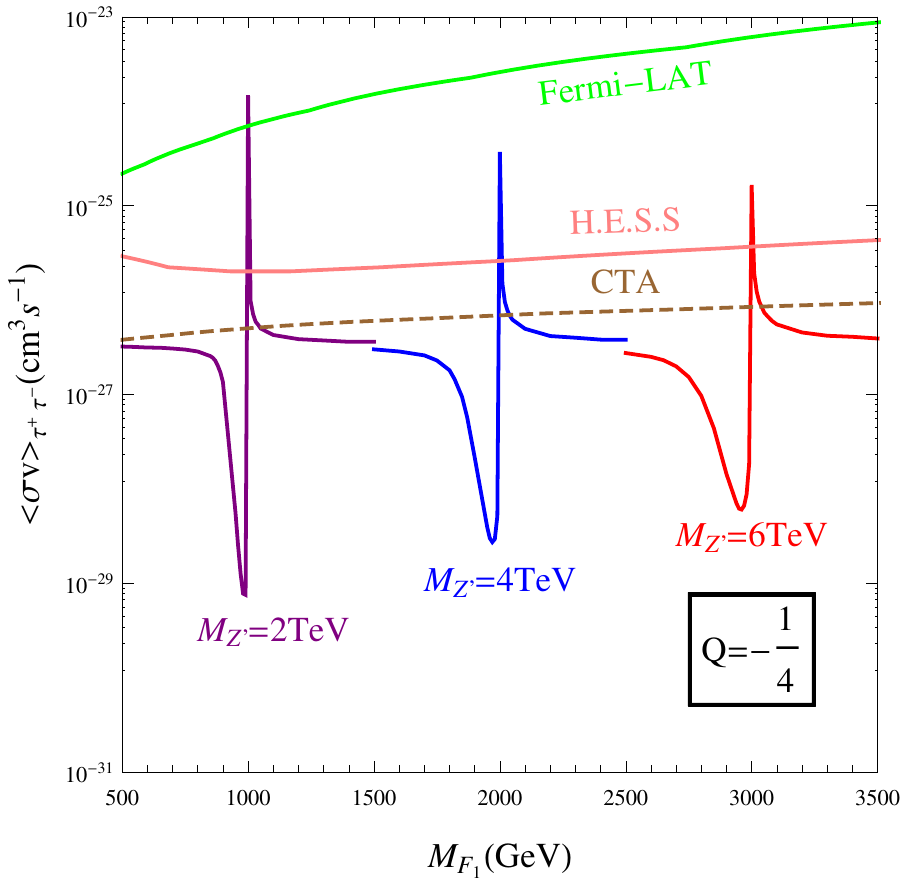}
\end{center}
\caption{Same as Fig.~\ref{Fig:IDS}, but for fermion DM $F_1$.
\label{Fig:IDF}}
\end{figure}

Now, we consider fermion DM. The results are shown in Fig.~\ref{Fig:IDF}. In spite of the cross sections are much larger than scalar DM, the effects of varying $B-L$ charge $Q$ and $M_{Z'}$ are similar. In left panel of Fig.~\ref{Fig:IDF}, we show the benchmark scenario $M_{F_1}=1~\TeV$ with $Q=-\frac{1}{4},-4,\frac{3}{2}$. Close to $M_{H_1^0}=M_{Z'}/2$, the maximum values of $\langle \sigma v \rangle_{\tau^+\tau^-}$ for $Q=-\frac{1}{4},-4$ are enhanced to the order of $10^{-24}\text{cm}^3\text{s}^{-1}$, which has already exceed Fermi-LAT limit. The maximum value for $Q=\frac{3}{2}$ is one order of magnitude smaller. Although this is lower than Fermi-LAT limit, it is still larger than H.E.S.S. limit. Future CTA limit will push the limit down to about $5\times10^{-27}\text{cm}^3\text{s}^{-1}$, hence large part of the resonance regions are detectable.
Moreover, when $M_{F_1}$ is slightly lighter than $M_{Z'}/2$, the cross sections dramatically  diminish to $3\times10^{-29}\text{cm}^3\text{s}^{-1}$, $8\times10^{-30}\text{cm}^3\text{s}^{-1}$,$6\times10^{-31}\text{cm}^3\text{s}^{-1}$ for $Q=\frac{3}{2},-\frac{1}{4},-4$ respectively. Therefore, this region can easily escape indirect detection limit. Right panel of Fig.~\ref{Fig:IDF} shows $\langle\sigma v \rangle_{\tau^+\tau^-}$ as a function of $M_{H_1^0}$, while we fix $Q=-\frac{1}{4}$ and vary $M_{Z'}=2,4,6~\TeV$. One can see that only $M_{Z'}=2~\TeV$ could exceed Fermi-LAT limit, since as $M_{H_1^0}(M_{Z'})$ getting bigger, the Fermi-LAT limit is less stringent while the enhancement effect is weaker. With clearly enhanced cross section, the resonance regions are all within the reach of H.E.S.S. and CTA.

In a nutshell, constraints from indirect detection can be neglected for scalar DM, even taking into account the Breit-Wigner enhancement effect. Meanwhile, for fermion DM, the indirect detection experiments are sensitive to the region close to $M_{F_1}\gtrsim M_{Z'}/2$, and the region $M_{F_1}\lesssim M_{Z'}/2$ can avoid indirect detection limit.

\section{Results}\label{Sec:CA}

In previous Sec.~\ref{Sec:LHC}-\ref{Sec:DM}, we perform a separated analysis on certain observables. Now, in this section, we first combine the separated analysis result from collider signature, dark radiation, DM relic density, direct and indirect detection for some benchmark scenarios. Based on the benchmark results, a scanning of the resonance region $M_\text{DM}\simeq M_{Z'}/2$ is then carried out.

\subsection{Combined Results}

\begin{figure}
\begin{center}
\includegraphics[width=0.45\linewidth]{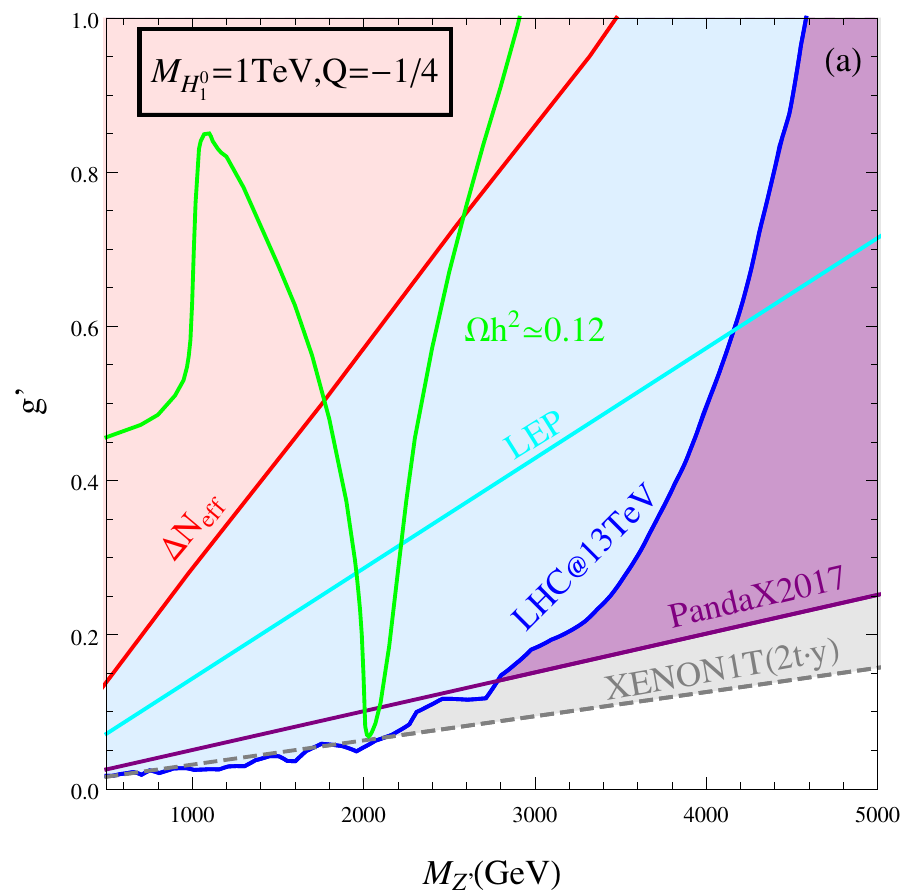}
\includegraphics[width=0.45\linewidth]{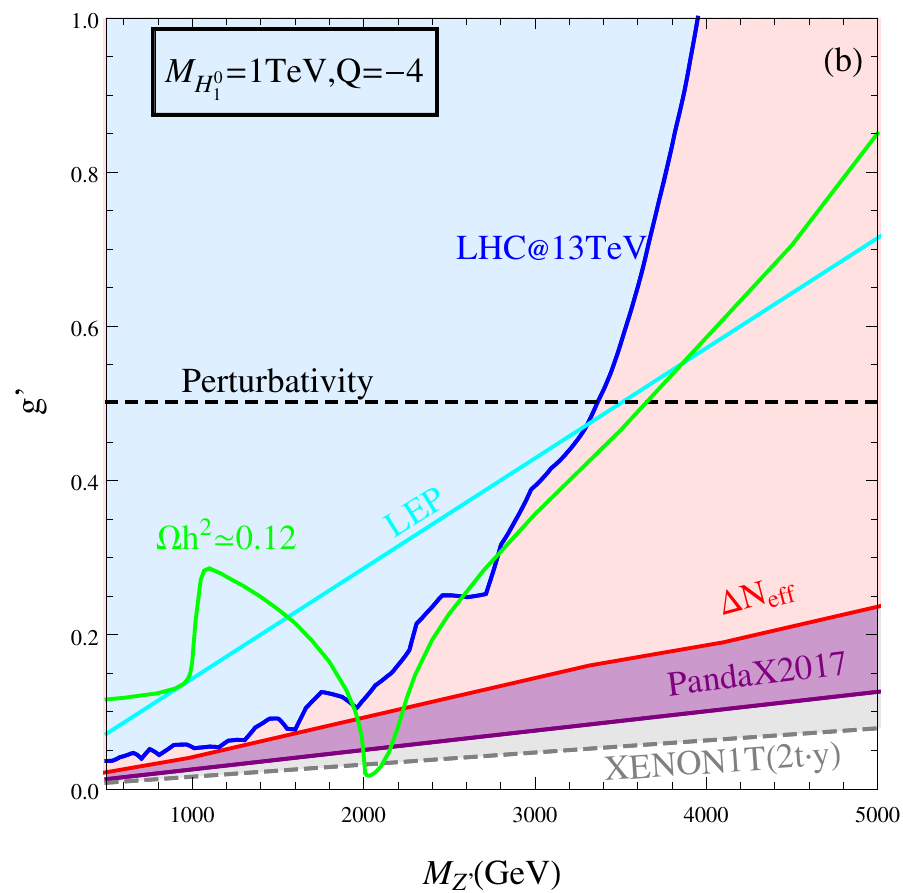}
\includegraphics[width=0.45\linewidth]{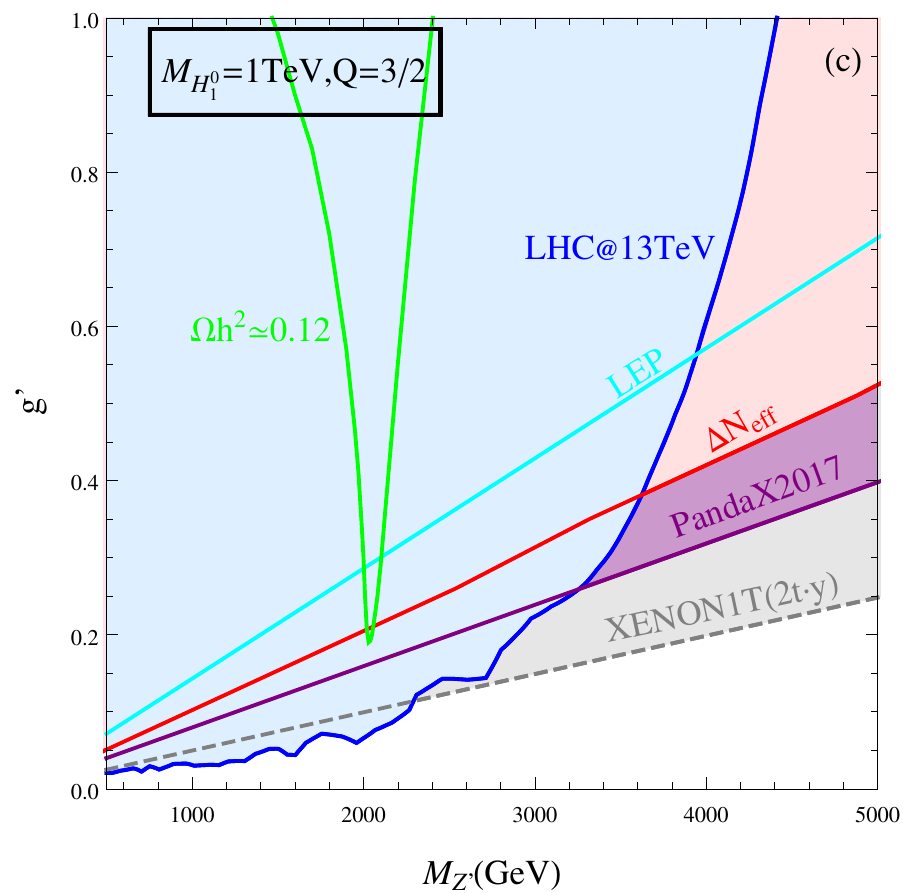}
\includegraphics[width=0.45\linewidth]{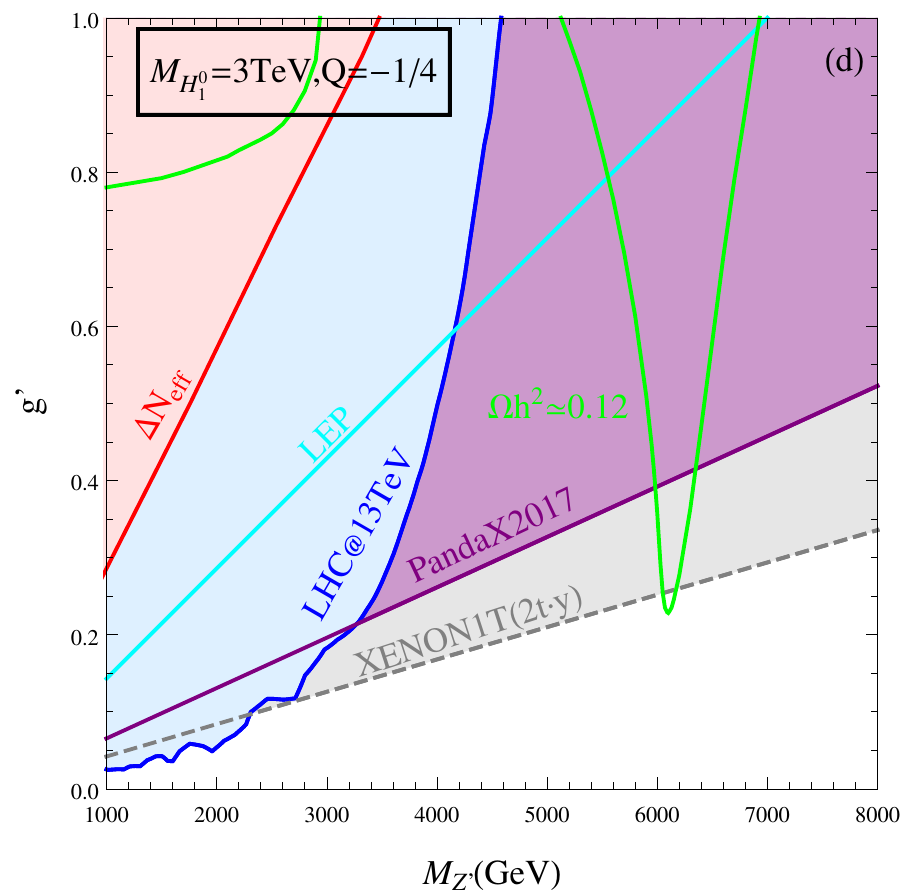}
\end{center}
\caption{Combined results in the $g'$-$M_{Z'}$ plane for scalar DM $M_{H_1^0}$. (a) $M_{H_1^0}=1~\TeV$ with $Q=-\frac{1}{4}$; (b) $M_{H_1^0}=1~\TeV$ with $Q=-4$; (c) $M_{H_1^0}=1~\TeV$ with $Q=\frac{3}{2}$; (d) $M_{H_1^0}=3~\TeV$ with $Q=-\frac{1}{4}$. Experimental bounds are obtained from 13 TeV LHC dilepton signature(blue)\cite{Aaboud:2017buh}, LEP precise measurement(cyan)\cite{Cacciapaglia:2006pk}, dark radiation observable $\Delta N_\text{eff}$(red)\cite{Ade:2015xua}, direct detection limit from PandaX2017(purple)\cite{Cui:2017nnn} and future XENON1T(2t$\cdot$y)(gray)\cite{Aprile:2015uzo}. With fixed DM mass and $B-L$ charge $Q$, the green lines further satisfy relic density condition $\Omega h^2\simeq0.12$ \cite{Ade:2015xua}. To make sure the model is perturbative, we also require $|g'\cdot Q_\text{max}|<\sqrt{2\pi}$(black).
\label{Fig:CBS}}
\end{figure}

Some general comments before we present the combined results. First, the LEP  precise measurement, dark radiation observable $\Delta N_\text{eff}$ and direct detection all set limits on $M_{Z'}/g'$, but there dependence on $B-L$ charge $Q$ is different. The benchmark scenarios $Q=-\frac{1}{4},-4,\frac{3}{2}$ with $M_\text{DM}=1~\TeV$ will show this.  Second, as already shown in Fig.~\ref{Fig:LHCex}, the LHC@13TeV dilepton signature is sensitive to $M_{Z'}\lesssim4~\TeV$. So for DM heavier than 2 TeV, there will be no constraint from LHC@13TeV. We choose one benchmark scenario $M_\text{DM}=3~\TeV$ with $Q=-\frac{1}{4}$ to illustrate this. Third, the perturbation condition $|g'\cdot Q_\text{max}|<\sqrt{2\pi}$ is also needed to keep the model perturbative. For $Q=-4$, $|Q_\text{max}|=|Q_\chi|=5$, so $g'<\sqrt{2\pi}/Q_\text{max}\approx0.5$ is required. For the other two choices $Q=-\frac{1}{4},\frac{3}{2}$, $g'<1$ will satisfy the perturbation condition.

Choosing four benchmark scenarios, the combined results for scalar DM $H_1^0$ are depicted in Fig.~\ref{Fig:CBS}. As discussed in Sec.~\ref{Sec:ID}, no parameter space for scalar DM is within the reach of indirect detection, so corresponding limits are missing in these figures. The benchmark scenario $M_{H_1^0}=1~\TeV$ with $Q=-\frac{1}{4}$ is shown in Fig.~\ref{Fig:CBS} (a). In this scenario, it is clear that nowadays the most stringent constraint is from LHC@13TeV when $M_{Z'}\lesssim2800~\GeV$ and it has already excluded all parameter space. Therefore, no signature is expected to be observed at ongoing direct detection experiments. The benchmark scenario $M_{H_1^0}=1~\TeV$ with $Q=-4$ is shown in Fig.~\ref{Fig:CBS} (b). In this scenario, constraints from $\Delta N_\text{eff}$ and direct detection are already severer than LHC@13TeV even in the low mass region. And only the resonance region $M_{Z'}\sim2~\TeV$ could avoid all constraints. The result for $M_{H_1^0}=1~\TeV$ with $Q=\frac{3}{2}$ is shown in Fig.~\ref{Fig:CBS} (c). In this case, although the $\Delta N_\text{eff}$ limit is severer than LEP, it is still weaker than PandaX2017. And the parameter space is again fully excluded by PandaX2017 and LHC@13TeV. The benchmark scenario for heavier DM $M_{H_1^0}=3~\TeV$ with $Q=-\frac{1}{4}$ is shown in Fig.~\ref{Fig:CBS} (d). With less stringent limits from direct detection, now the resonance region can escape PandaX2017 limit, but most region are still in the reach of XENON1T(2t$\cdot$y).

\begin{figure}
\begin{center}
\includegraphics[width=0.45\linewidth]{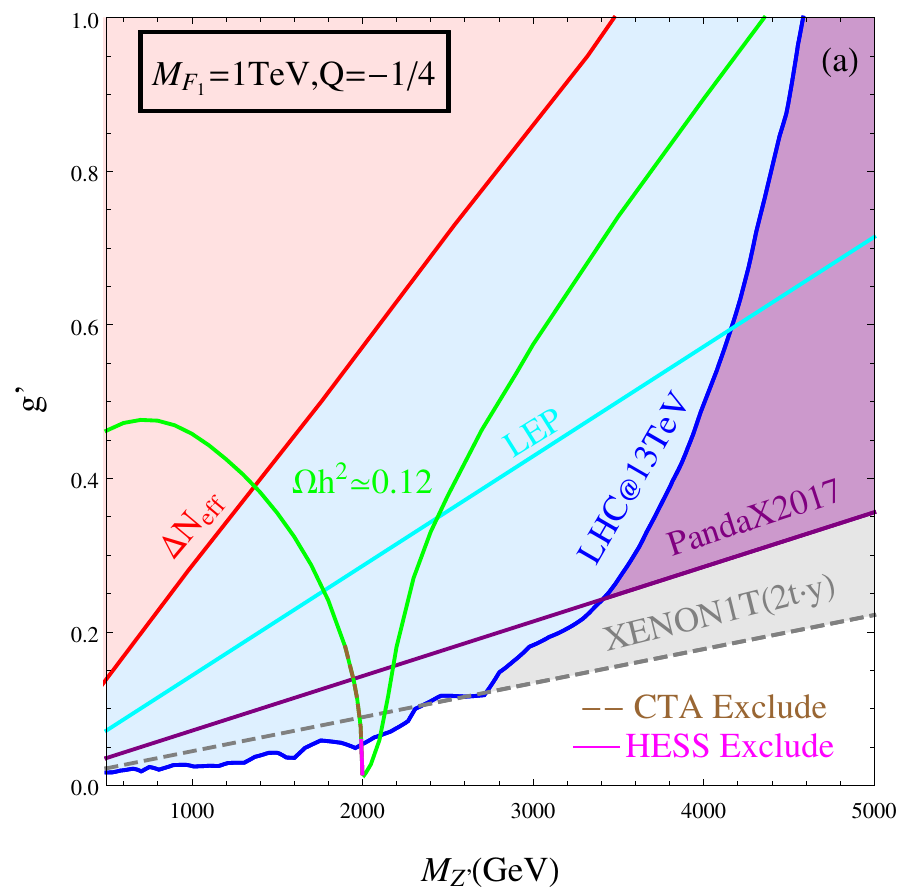}
\includegraphics[width=0.45\linewidth]{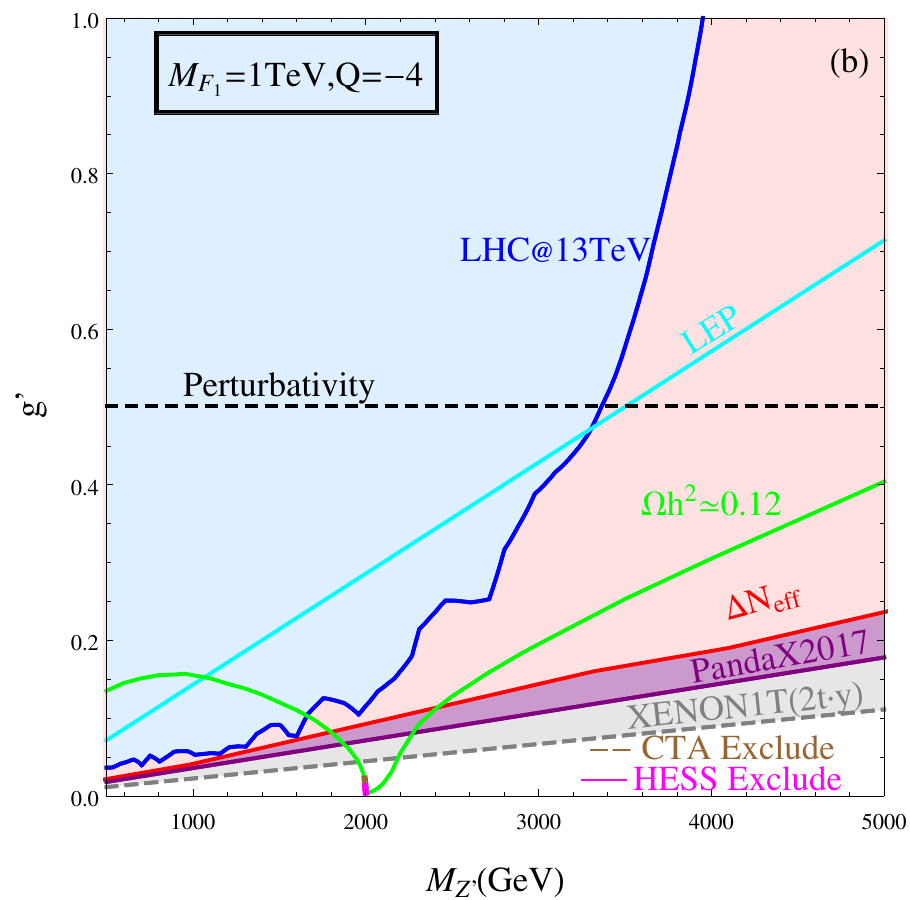}
\includegraphics[width=0.45\linewidth]{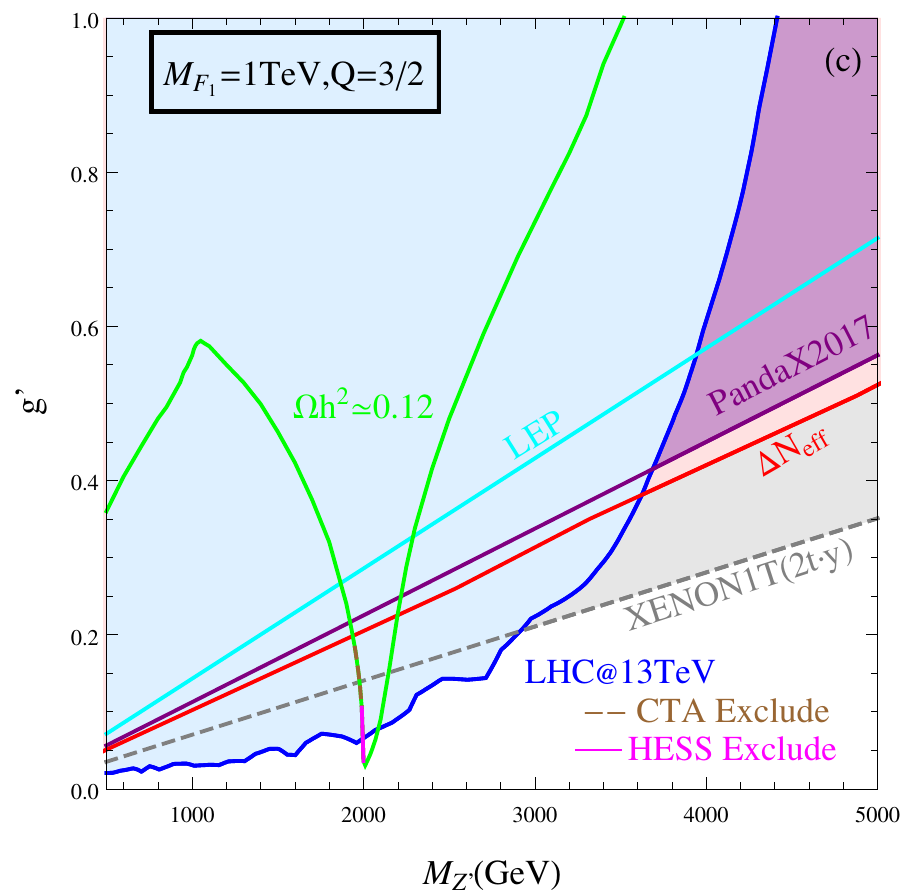}
\includegraphics[width=0.45\linewidth]{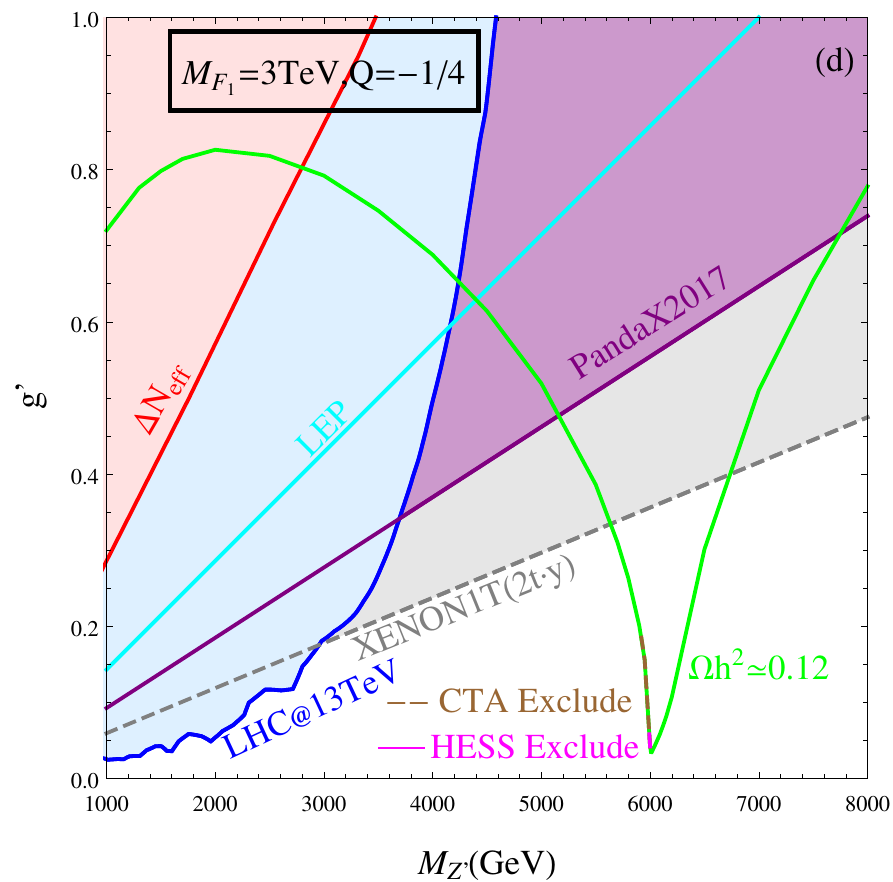}
\end{center}
\caption{Same as Fig.~\ref{Fig:CBS}, but for fermion DM $F_1$. In addition, the pink and brown curves are excluded by H.E.S.S.\cite{Abdallah:2016ygi} and future CTA \cite{Lefranc:2015pza} indirect detection experiments.
\label{Fig:CBF}}
\end{figure}

Then, the combined results for fermion DM $F_1$ is shown in Fig.~\ref{Fig:CBF}. Comparing with scalar DM, the indirect detection experiments could also exclude certain parameter space. Instead of deriving the whole exclusion region in the $g'$-$M_{Z'}$ plane, we only get the excluded curves satisfying relic density condition, which can be easily obtained from Fig.~\ref{Fig:IDF} (a). The benchmark scenario $M_{F_1}=1~\TeV$ with $Q=-\frac{1}{4}$ is shown in Fig.~\ref{Fig:CBF} (a). It is amazing that the combined result of LHC@13TeV and H.E.S.S. has already excluded  $M_{F_1}\gtrsim M_{Z'}/2$. So neither direct detection  nor indirect detection experiment will observe positive signature. And the only way to probe the allowed region $M_{F_1}\lesssim M_{Z'}/2$ is via dilepton signature at ongoing LHC. Similar arguments are also true for benchmark scenario  $M_{F_1}=1~\TeV$ with $Q=\frac{3}{2}$ as shown in Fig.~\ref{Fig:CBF} (c). Note in this scenario, the constraint from $\Delta N_\text{eff}$ is more stringent than current direct detection PandaX2017. The result for $M_{F_1}=1~\TeV$ with $Q=-4$ is shown in Fig.~\ref{Fig:CBF} (b). In this scenario, the $M_{F_1}\gtrsim M_{Z'}/2$ region, i.e. $1880~\GeV\lesssim M_{F_1}\lesssim1995 ~\GeV$, is also viable. And either direct detection or indirect detection experiment is hopefully to have some positive signature. Fig.~\ref{Fig:CBF} (d) shows the result of $M_{F_1}=3~\TeV$ with $Q=-\frac{1}{4}$. In this scenario, the constraints are relative loose. Hence, one can see that $M_{Z'}$ can be over about 1 TeV away from the resonance nowadays. Again similar as Fig.~\ref{Fig:CBF} (b), no region could be cross checked by direct and indirect detection.

\subsection{Scanning Results}

\begin{figure}
\begin{center}
\includegraphics[width=0.45\linewidth]{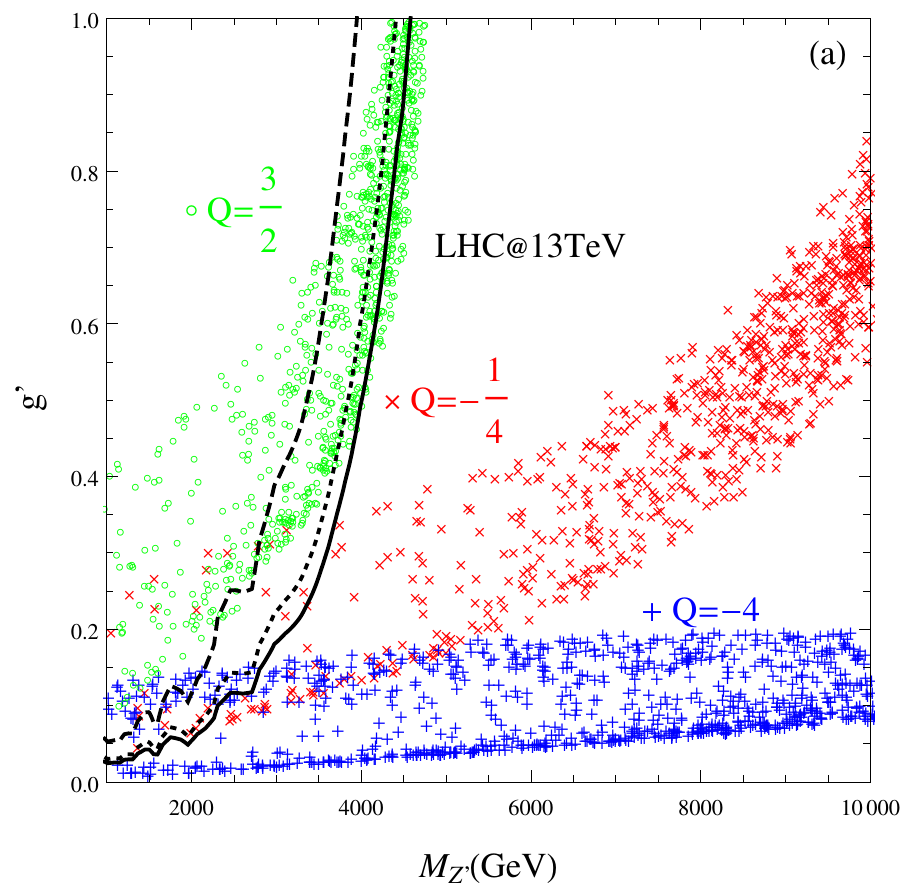}
\includegraphics[width=0.45\linewidth]{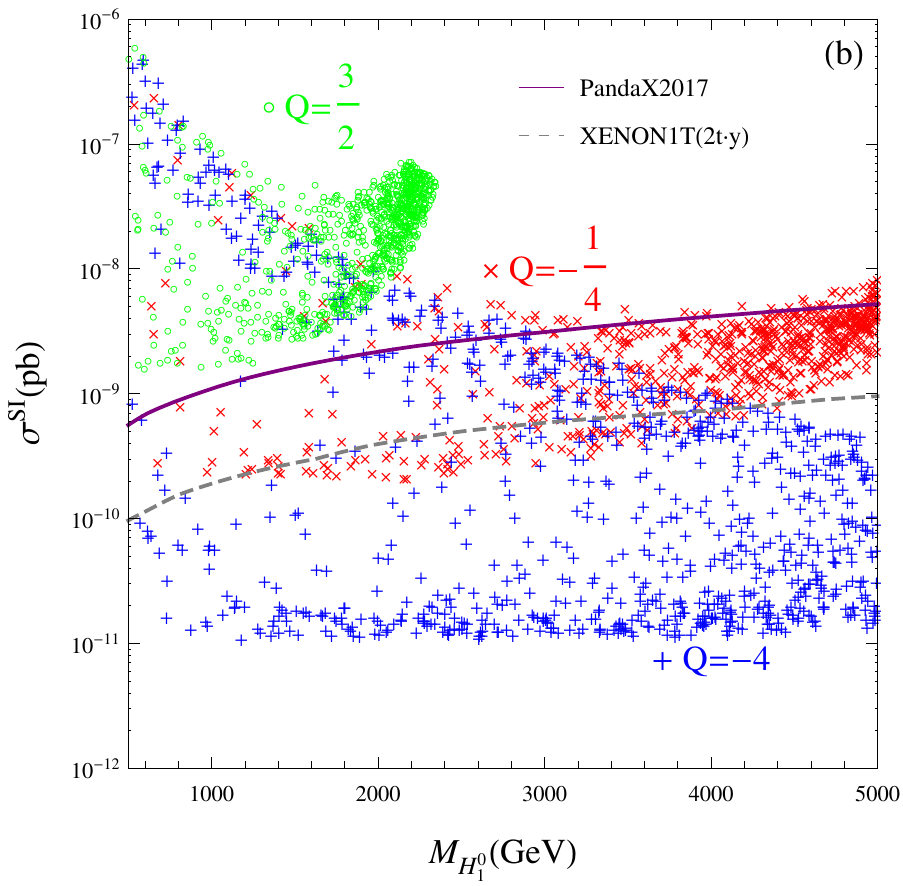}
\includegraphics[width=0.45\linewidth]{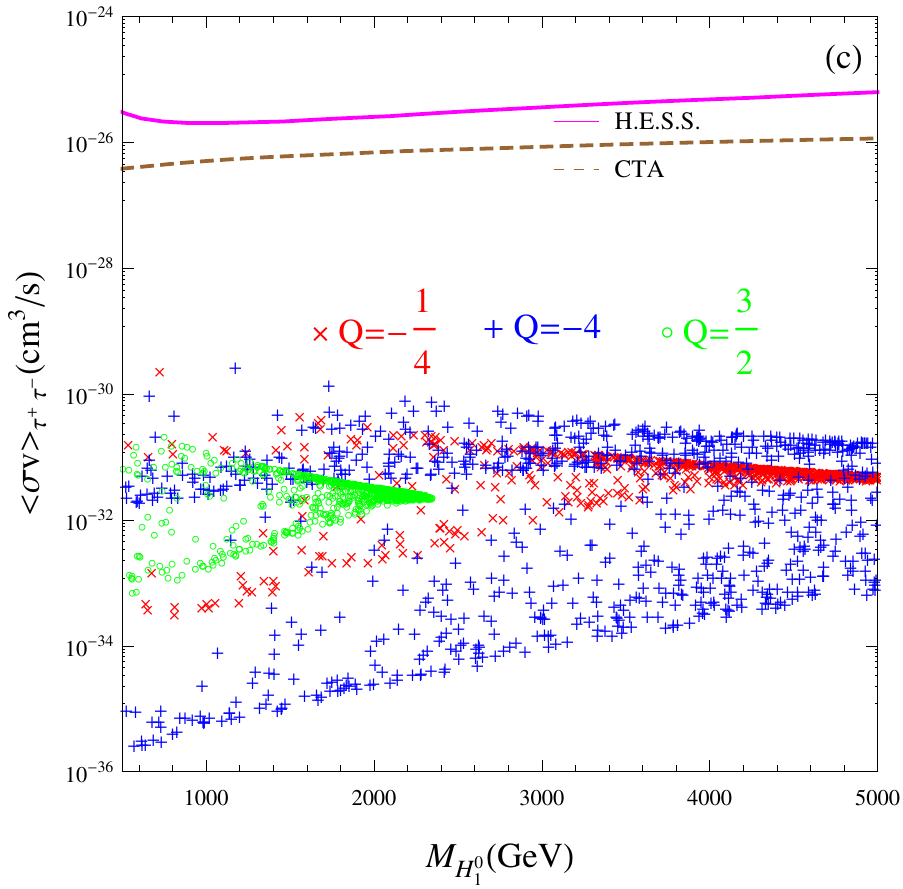}
\includegraphics[width=0.45\linewidth]{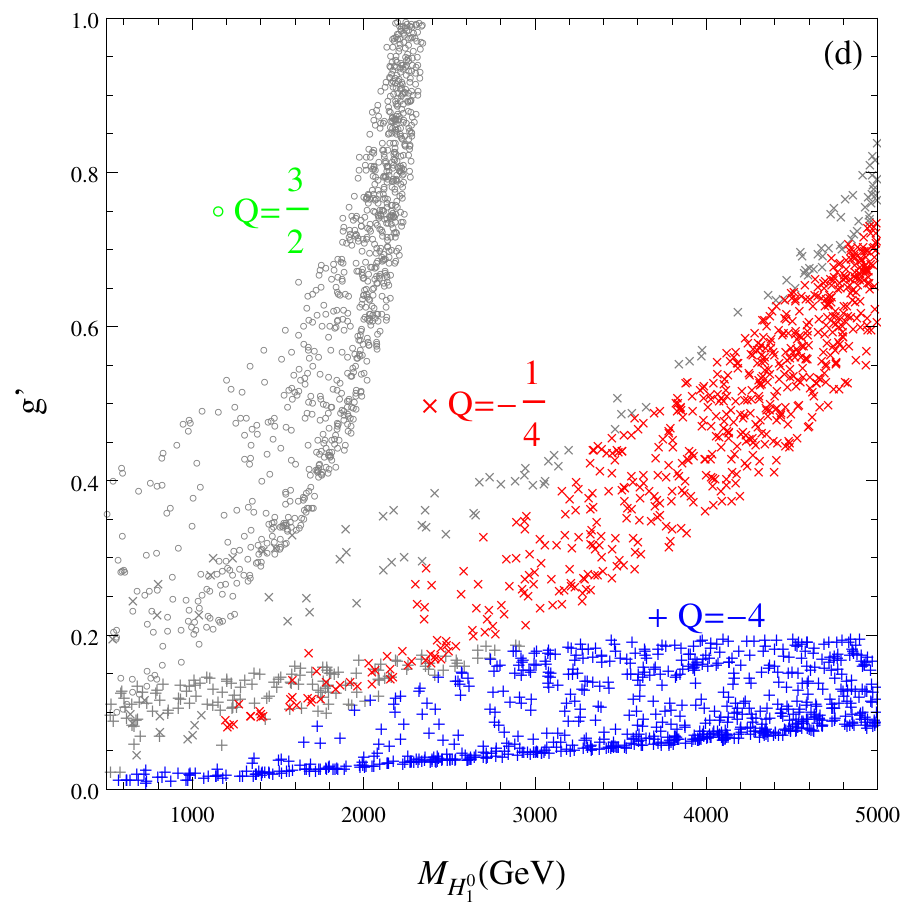}
\end{center}
\caption{Scanning results for scalar DM $H_1^0$. (a) Viable parameter space for relic density in the $g'$-$M_{Z'}$ plane. The solid, dotted and dashed lines (from the bottom up) correspond to LHC@13TeV exclusion limits for $Q=-\frac{1}{4},\frac{3}{2}$ and ${-4}$ respectively. (b) Spin-independent cross section as a function of $M_{H_1^0}$. (c) Annihilation cross section into $\tau^+\tau^-$ final states as a function of $M_{H_1^0}$. (d) Allowed parameter space after applying all constraints. The gray points are excluded.
\label{Fig:SCS}}
\end{figure}

Based on the combined results of benchmark scenario, one is aware that the resonance region $M_\text{DM}\simeq M_{Z'}/2$ is the viable parameter space to satisfy all experimental constraints.
Hence, instead of fully scan the whole parameter space, we focus on the resonance region $M_\text{DM}\simeq M_{Z'}/2$. We fix $Q=-\frac{1}{4},-4$ or $\frac{3}{2}$, and then scan in the
following range
\begin{eqnarray}
g'\in[0,1], M_\text{DM}\in[500,5000]~\GeV, M_{Z'}-2M_\text{DM}\in[-50,100]~\GeV
\end{eqnarray}
for both scalar and fermion DM. For scalar DM, the direct detection could exclude the samples out of the mass interval $M_{Z'}-2M_{H_1^0}\in[-50,100]~\GeV$. In principle, the mass interval $|M_{Z'}-2M_{F_1}|$ for fermion DM could be $\mathcal{O}(\TeV)$, but the indirect detection experiments are only sensitive to the near resonance region. We focus on this region so as well to probe the less considered indirect signatures. During the scan, we require the relic density satisfies Planck result in $3\sigma$ range, i.e., $0.1118<\Omega h^2<0.1280$. For each case, we generate over $5\times10^4$ samples, and get about $1\times10^3$ survived samples. Constraints from LHC dilepton signature, direct and indirect detection are further taken into account to acquire the allowed parameter space at present, because these provide the most stringent constraints under certain circumstances according to the combined results \footnote{Note that for fermion DM $F_1$ with $Q=\frac{3}{2}$, one should also take into account the constraint from $\Delta N_\text{eff}$, since it could be more stringent than direct detection.}.

The scanning results for scalar DM $H_1^0$ are shown in Fig.~\ref{Fig:SCS}. Fig.~\ref{Fig:SCS} (a) presents the result in the $g'$-$M_{Z'}$ plane. Basically speaking, $g'$ increases as $M_{Z'}$ increases. For $Q=\frac{3}{2}$, $g'$ is larger than 1 when $M_{Z'}$ is larger than about $4~\TeV$, and most samples are already excluded by LHC@13TeV. For $Q=-\frac{1}{4}$, LHC@13TeV could fully exclude $M_{Z'}\lesssim2~\TeV$, while for $Q=-4$, $M_{Z'}$ downs to about 1 TeV is still possible. Fig.~\ref{Fig:SCS} (b) shows the predicted spin-independent DM-nucleon scattering cross section as a function of $M_{H_1^0}$. Clearly, the case of $Q=\frac{3}{2}$ is fully excluded by PandaX2017. Although for $Q=-\frac{1}{4}$, most samples could survive under PandaX2017, most samples are within the reach of XENON1T(2t$\cdot$y). And if no signature is observed by XENON1T(2t$\cdot$y), $M_{H_1^0}$ should fall in the range $[1200,4400]~\GeV$. As for $Q=-4$, most samples could escape direct detection limits even in the future. The annihilation cross section into $\tau^+\tau^-$ final states is also presented in Fig.~\ref{Fig:SCS} (c) for completeness, even though it is serval orders of magnitude under future CTA limit. The survived simples after imposing all constraints are depicted in Fig.~\ref{Fig:SCS} (d). For $Q=\frac{3}{2}$, no survived points are obtained, due to tight direct detection constraints. For $Q=-\frac{1}{4}$, one needs $M_{H_1^0}\gtrsim1.2~\TeV$ to escape all constraints, while for $Q=-4$, sub-TeV $H_1^0$ with small $g'\sim0.01$ is still viable.

\begin{figure}
\begin{center}
\includegraphics[width=0.45\linewidth]{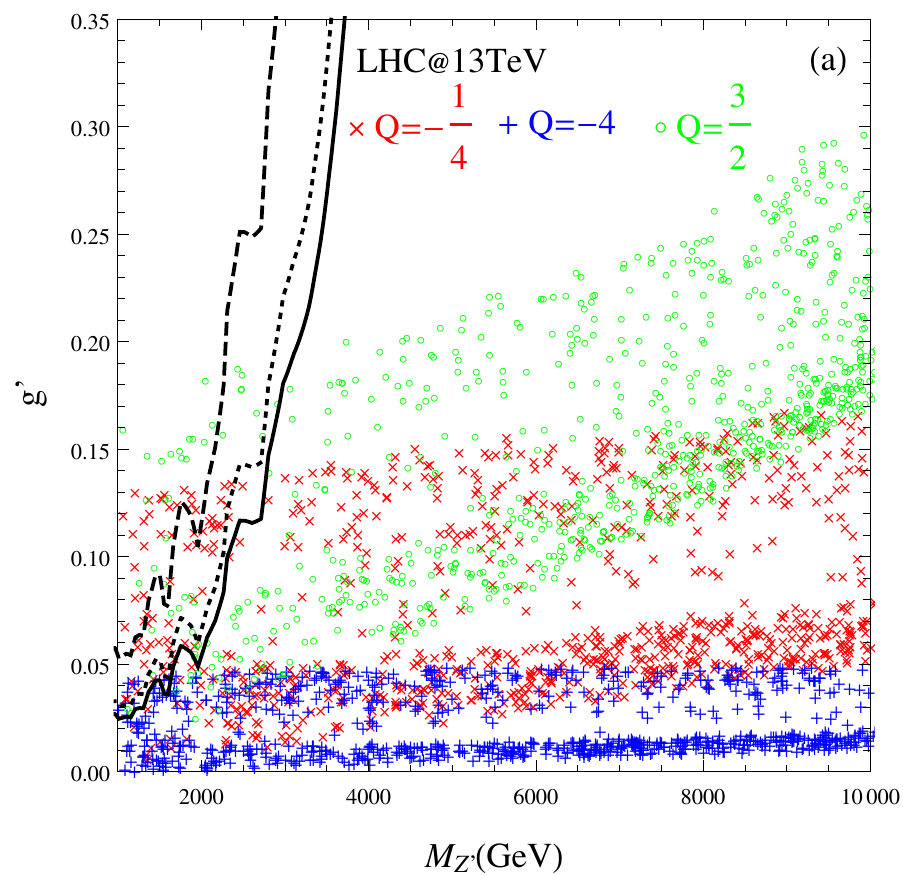}
\includegraphics[width=0.45\linewidth]{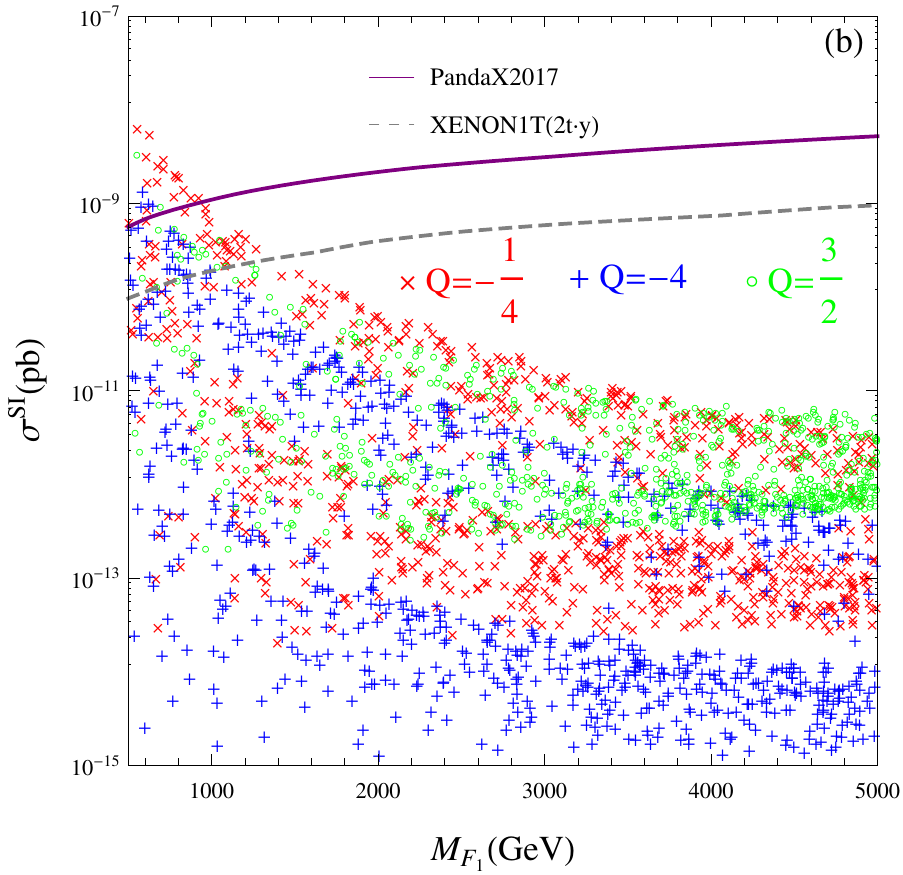}
\includegraphics[width=0.45\linewidth]{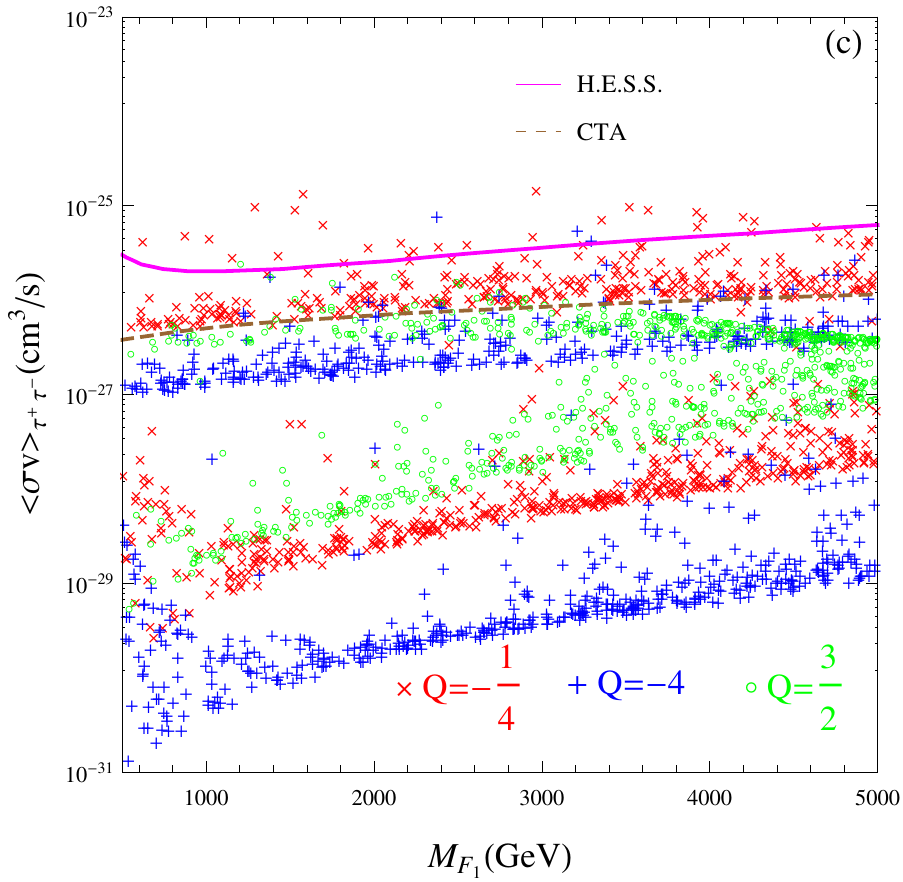}
\includegraphics[width=0.45\linewidth]{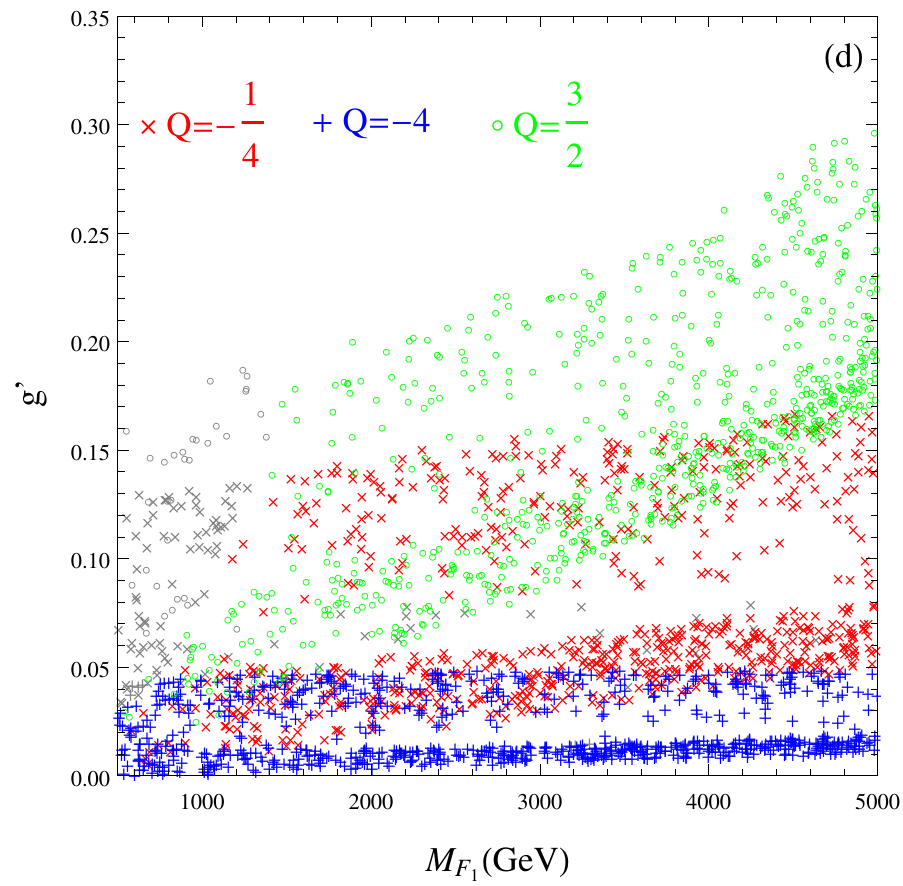}
\end{center}
\caption{Same as Fig.~\ref{Fig:SCS}, but for fermion DM $F_1$.
\label{Fig:SCF}}
\end{figure}

The scanning results for fermion DM $F_1$ are shown in Fig.~\ref{Fig:SCF}. Viable parameter space for relic density in the $g'$-$M_{Z'}$ plane are shown in Fig.~\ref{Fig:SCF}~(a). Comparing with scalar DM, a smaller $g'$ is usually needed for same value of $M_\text{DM}$ and $Q$. Hence, it is more easier for fermion DM to satisfy the constraints from LHC@13TeV. Especially for $Q=-4$, no sample is excluded by LHC@13TeV. According to direct detection results in Fig.~\ref{Fig:SCF}~(b), only small amount of samples with $M_{F_1}\lesssim1~\TeV$ are excluded. As for indirect detection shown in Fig.~\ref{Fig:SCF}~(c), the most promising case is $Q=-\frac{1}{4}$. With few samples excluded by H.E.S.S., about half of the samples are within reach of future CTA experiment for $Q=-\frac{1}{4}$. For the other two cases, a little of the samples could lead to observable indirect signatures even in the future. Samples passing through all constraints are shown in Fig.~\ref{Fig:SCF}~(d). It is clear that most samples are survived for all three choices of $Q$.

\section{Conclusions}\label{Sec:CL}

In this paper, the phenomenology of $Z'$ portal scalar and Dirac fermion DM in $B-L$ scotogenic Dirac model is comprehensively studied. With unconventional $B-L$ charge $Q$ of right-handed neutrino $\nu_R$, the Dirac neutrino mass induced at one-loop level, DM stability and anomaly cancellations are naturally realised. Depending on the mass spectrum, both scalar DM $H_1^0$ with $B-L$ charge $Q_{H_1^0}\simeq Q-1$ and fermion DM $F_1$ with $Q_{F_{1L}}=1$, $Q_{F_{1R}}=-Q$ are allowed. Three typical values $Q=-\frac{1}{4},-4,\frac{3}{2}$ are chosen to illustrate the testability of this model.

Phenomenology aspects are detail examined in collider signature, dark radiation, DM relic density, direct and indirect detection. Generally speaking, constraints from dilepton search at 13 TeV LHC usually place the most stringent exclusion limits in the lower mass region, while direct detection limits becomes the most stringent in the higher mass region. Combining all the constraints leads to the viable parameter space falling in the resonance region $M_\text{DM}\sim M_{Z'}/2$. In addition, indirect detection could also exclude the region very close to the resonance region $M_{F_1}\gtrsim M_{Z'}/2$ for Dirac fermion DM. Notably, direct detection is sensitive to the region away from the resonance region, so next generation direct and indirect detection experiment might hard to both have positive signatures provided Dirac fermion DM. Meanwhile, this model also predict testable $\Delta N_\text{eff}$ for TeV-scale $Z'$, which is an intrinsic character that can be used to distinguish this Dirac model from the Majorona model for neutrino mass.

Scanning on the resonance region, allowed parameter space is then obtained. As for scalar DM, the case of $Q=\frac{3}{2}$ is already fully excluded by PandaX2017, while for $Q=-\frac{1}{4}$, most samples allowed at present are within the reach of future XENON1T(2t$\cdot$y). For $Q=-4$, sub-TeV DM is still viable. Comparing with scalar DM, the Dirac fermion DM is easier to satisfy all constraints. For $Q=-\frac{1}{4}$, it is promising at future indirect detection CTA experiment, meanwhile it is less promising for $Q=-4$ and $\frac{3}{2}$.

\section*{Acknowledgments}

The work of Weijian Wang is supported by National Natural Science
Foundation of China under Grant Numbers 11505062, Special Fund of
Theoretical Physics under Grant Numbers 11447117 and Fundamental
Research Funds for the Central Universities under Grant Numbers
 2016MS133.


\begin{thebibliography}{000}

\bibitem{Weinberg:1979sa}
  S.~Weinberg,
  Phys.\ Rev.\ Lett.\  {\bf 43}, 1566 (1979).

\bibitem{Gu:2006dc}
  P.~H.~Gu and H.~J.~He,
  JCAP {\bf 0612}, 010 (2006)
  [hep-ph/0610275].

\bibitem{Gu:2007ug}
  P.~H.~Gu and U.~Sarkar,
  Phys.\ Rev.\ D {\bf 77}, 105031 (2008)
  [arXiv:0712.2933 [hep-ph]].

\bibitem{Farzan:2012sa}
  Y.~Farzan and E.~Ma,
  Phys.\ Rev.\ D {\bf 86}, 033007 (2012)
  [arXiv:1204.4890 [hep-ph]].

\bibitem{Chulia:2016ngi}
  S.~Centelles Chulia, E.~Ma, R.~Srivastava and J.~W.~F.~Valle,
  Phys.\ Lett.\ B {\bf 767}, 209 (2017)
  [arXiv:1606.04543 [hep-ph]].

\bibitem{Bonilla:2016diq}
  C.~Bonilla, E.~Ma, E.~Peinado and J.~W.~F.~Valle,
  Phys.\ Lett.\ B {\bf 762}, 214 (2016)
  [arXiv:1607.03931 [hep-ph]].

\bibitem{Ma:2016mwh}
  E.~Ma and O.~Popov,
  Phys.\ Lett.\ B {\bf 764}, 142 (2017)
  [arXiv:1609.02538 [hep-ph]].

\bibitem{Wang:2016lve}
  W.~Wang and Z.~L.~Han,
  JHEP {\bf 1704}, 166 (2017)
  [arXiv:1611.03240 [hep-ph]].

\bibitem{Borah:2017leo}
  D.~Borah and A.~Dasgupta,
  JCAP {\bf 1706}, no. 06, 003 (2017)
  [arXiv:1702.02877 [hep-ph]].

\bibitem{Wang:2017mcy}
  W.~Wang, R.~Wang, Z.~L.~Han and J.~Z.~Han,
  Eur.\ Phys.\ J.\ C {\bf 77}, no. 12, 889 (2017)
  [arXiv:1705.00414 [hep-ph]].

\bibitem{CentellesChulia:2017koy}
  S.~Centelles Chulia, R.~Srivastava and J.~W.~F.~Valle,
  Phys.\ Lett.\ B {\bf 773}, 26 (2017)
  [arXiv:1706.00210 [hep-ph]].

\bibitem{Ma:2017kgb}
  E.~Ma and U.~Sarkar,
  Phys.\ Lett.\ B {\bf 776}, 54 (2018)
  [arXiv:1707.07698 [hep-ph]].

\bibitem{Yao:2017vtm}
  C.~Y.~Yao and G.~J.~Ding,
  Phys.\ Rev.\ D {\bf 96}, no. 9, 095004 (2017)
  [arXiv:1707.09786 [hep-ph]].

\bibitem{Bonilla:2017ekt}
  C.~Bonilla, J.~M.~Lamprea, E.~Peinado and J.~W.~F.~Valle,
  Phys.\ Lett.\ B {\bf 779}, 257 (2018)
  [arXiv:1710.06498 [hep-ph]].

\bibitem{Ibarra:2017tju}
  A.~Ibarra, A.~Kushwaha and S.~K.~Vempati,
  Phys.\ Lett.\ B {\bf 780}, 86 (2018)
  [arXiv:1711.02070 [hep-ph]].

\bibitem{Borah:2017dmk}
  D.~Borah and B.~Karmakar,
  Phys.\ Lett.\ B {\bf 780}, 461 (2018)
  [arXiv:1712.06407 [hep-ph]].

\bibitem{Yao:2018ekp}
  C.~Y.~Yao and G.~J.~Ding,
  arXiv:1802.05231 [hep-ph].

\bibitem{CentellesChulia:2018gwr}
  S.~Centelles Chulia, R.~Srivastava and J.~W.~F.~Valle,
  Phys.\ Lett.\ B {\bf 781}, 122 (2018)
  [arXiv:1802.05722 [hep-ph]].

\bibitem{CentellesChulia:2018bkz}
  S.~Centelles Chulia, R.~Srivastava and J.~W.~F.~Valle,
  arXiv:1804.03181 [hep-ph].

\bibitem{tHooft:1976rip}
  G.~'t Hooft,
  Phys.\ Rev.\ Lett.\  {\bf 37}, 8 (1976).

\bibitem{Kuzmin:1985mm}
  V.~A.~Kuzmin, V.~A.~Rubakov and M.~E.~Shaposhnikov,
  Phys.\ Lett.\  {\bf 155B}, 36 (1985).

\bibitem{Heeck:2013rpa}
  J.~Heeck and W.~Rodejohann,
  EPL {\bf 103}, no. 3, 32001 (2013)
  [arXiv:1306.0580 [hep-ph]].

\bibitem{Heeck:2013vha}
  J.~Heeck,
  Phys.\ Rev.\ D {\bf 88}, 076004 (2013)
  [arXiv:1307.2241 [hep-ph]].

\bibitem{Ma:2006km}
  E.~Ma,
  Phys.\ Rev.\ D {\bf 73}, 077301 (2006)
  [hep-ph/0601225].

\bibitem{Ding:2016wbd}
  R.~Ding, Z.~L.~Han, Y.~Liao and W.~P.~Xie,
  JHEP {\bf 1605}, 030 (2016)
  [arXiv:1601.06355 [hep-ph]].

\bibitem{Alves:2013tqa}
  A.~Alves, S.~Profumo and F.~S.~Queiroz,
  JHEP {\bf 1404}, 063 (2014)
  [arXiv:1312.5281 [hep-ph]].

\bibitem{Buchmueller:2014yoa}
  O.~Buchmueller, M.~J.~Dolan, S.~A.~Malik and C.~McCabe,
  JHEP {\bf 1501}, 037 (2015)
  [arXiv:1407.8257 [hep-ph]].

\bibitem{Alves:2015pea}
  A.~Alves, A.~Berlin, S.~Profumo and F.~S.~Queiroz,
  Phys.\ Rev.\ D {\bf 92}, no. 8, 083004 (2015)
  [arXiv:1501.03490 [hep-ph]].

\bibitem{Wang:2015saa}
  W.~Wang and Z.~L.~Han,
  Phys.\ Rev.\ D {\bf 92}, 095001 (2015)
  [arXiv:1508.00706 [hep-ph]].

\bibitem{Alves:2015mua}
  A.~Alves, A.~Berlin, S.~Profumo and F.~S.~Queiroz,
  JHEP {\bf 1510}, 076 (2015)
  [arXiv:1506.06767 [hep-ph]].

\bibitem{Jacques:2016dqz}
  T.~Jacques, A.~Katz, E.~Morgante, D.~Racco, M.~Rameez and A.~Riotto,
  JHEP {\bf 1610}, 071 (2016)
  [arXiv:1605.06513 [hep-ph]].

\bibitem{Fairbairn:2016iuf}
  M.~Fairbairn, J.~Heal, F.~Kahlhoefer and P.~Tunney,
  JHEP {\bf 1609}, 018 (2016)
  [arXiv:1605.07940 [hep-ph]].

\bibitem{Kaneta:2016vkq}
  K.~Kaneta, Z.~Kang and H.~S.~Lee,
  JHEP {\bf 1702}, 031 (2017)
  [arXiv:1606.09317 [hep-ph]].
  
\bibitem{Altmannshofer:2016jzy}
  W.~Altmannshofer, S.~Gori, S.~Profumo and F.~S.~Queiroz,
  JHEP {\bf 1612}, 106 (2016)
  [arXiv:1609.04026 [hep-ph]].

\bibitem{Singirala:2017see}
  S.~Singirala, R.~Mohanta and S.~Patra,
  arXiv:1704.01107 [hep-ph].

\bibitem{Ding:2017jdr}
  R.~Ding, Z.~L.~Han, L.~Feng and B.~Zhu,
  arXiv:1712.02021 [hep-ph].
  
\bibitem{Han:2017ars}
  Z.~L.~Han, W.~Wang and R.~Ding,
  Eur.\ Phys.\ J.\ C {\bf 78}, no. 3, 216 (2018)
  [arXiv:1712.05722 [hep-ph]].

\bibitem{Ma:2018bjw}
  E.~Ma,
  arXiv:1803.03891 [hep-ph].

\bibitem{Okada:2018ktp}
  S.~Okada,
  arXiv:1803.06793 [hep-ph].

\bibitem{FileviezPerez:2018toq}
  P.~Fileviez Perez and C.~Murgui,
  arXiv:1803.07462 [hep-ph].

\bibitem{Biswas:2018yus}
  A.~Biswas, S.~Choubey and S.~Khan,
  arXiv:1805.00568 [hep-ph].
  
\bibitem{Kanemura:2011vm}
  S.~Kanemura, O.~Seto and T.~Shimomura,
  Phys.\ Rev.\ D {\bf 84}, 016004 (2011)
  [arXiv:1101.5713 [hep-ph]].

\bibitem{Ma:2001kg}
  E.~Ma,
  Mod.\ Phys.\ Lett.\ A {\bf 17}, 535 (2002)
  [hep-ph/0112232].

\bibitem{Ding:2018jdk}
  R.~Ding, Z.~L.~Han, L.~Huang and Y.~Liao,
  arXiv:1802.05248 [hep-ph].

\bibitem{Aad:2012tfa}
  G.~Aad {\it et al.} [ATLAS Collaboration],
  Phys.\ Lett.\ B {\bf 716}, 1 (2012)
  [arXiv:1207.7214 [hep-ex]].

\bibitem{Chatrchyan:2012xdj}
  S.~Chatrchyan {\it et al.} [CMS Collaboration],
  Phys.\ Lett.\ B {\bf 716}, 30 (2012)
  [arXiv:1207.7235 [hep-ex]].

\bibitem{Robens:2015gla}
  T.~Robens and T.~Stefaniak,
  Eur.\ Phys.\ J.\ C {\bf 75}, 104 (2015)
  [arXiv:1501.02234 [hep-ph]].

\bibitem{Cacciapaglia:2006pk}
  G.~Cacciapaglia, C.~Csaki, G.~Marandella and A.~Strumia,
  Phys.\ Rev.\ D {\bf 74}, 033011 (2006)
  [hep-ph/0604111].

\bibitem{Basso:2008iv}
  L.~Basso, A.~Belyaev, S.~Moretti and C.~H.~Shepherd-Themistocleous,
  Phys.\ Rev.\ D {\bf 80}, 055030 (2009)
  [arXiv:0812.4313 [hep-ph]].
  L.~Basso, A.~Belyaev, S.~Moretti, G.~M.~Pruna and C.~H.~Shepherd-Themistocleous,
  Eur.\ Phys.\ J.\ C {\bf 71}, 1613 (2011)
  [arXiv:1002.3586 [hep-ph]].

\bibitem{Aad:2014cka}
  G.~Aad {\it et al.} [ATLAS Collaboration],
  Phys.\ Rev.\ D {\bf 90}, no. 5, 052005 (2014)
  [arXiv:1405.4123 [hep-ex]].
  M.~Aaboud {\it et al.} [ATLAS Collaboration],
  Phys.\ Lett.\ B {\bf 761}, 372 (2016)
  [arXiv:1607.03669 [hep-ex]].

\bibitem{Khachatryan:2016zqb}
  V.~Khachatryan {\it et al.} [CMS Collaboration],
  Phys.\ Lett.\ B {\bf 768}, 57 (2017)
  [arXiv:1609.05391 [hep-ex]].
  A.~M.~Sirunyan {\it et al.} [CMS Collaboration],
  arXiv:1803.06292 [hep-ex].

\bibitem{Aaboud:2017buh}
  M.~Aaboud {\it et al.} [ATLAS Collaboration],
  JHEP {\bf 1710}, 182 (2017)
  [arXiv:1707.02424 [hep-ex]].

\bibitem{Okada:2016gsh}
  N.~Okada and S.~Okada,
  Phys.\ Rev.\ D {\bf 93}, no. 7, 075003 (2016)
  [arXiv:1601.07526 [hep-ph]].

\bibitem{Klasen:2016qux}
  M.~Klasen, F.~Lyonnet and F.~S.~Queiroz,
  Eur.\ Phys.\ J.\ C {\bf 77}, no. 5, 348 (2017)
  [arXiv:1607.06468 [hep-ph]].

\bibitem{Okada:2016tci}
  N.~Okada and S.~Okada,
  Phys.\ Rev.\ D {\bf 95}, no. 3, 035025 (2017)
  [arXiv:1611.02672 [hep-ph]].

\bibitem{DeRomeri:2017oxa}
  V.~De Romeri, E.~Fernandez-Martinez, J.~Gehrlein, P.~A.~N.~Machado and V.~Niro,
  JHEP {\bf 1710}, 169 (2017)
  [arXiv:1707.08606 [hep-ph]].

\bibitem{Alloul:2013bka}
  A.~Alloul, N.~D.~Christensen, C.~Degrande, C.~Duhr and B.~Fuks,
  Comput.\ Phys.\ Commun.\  {\bf 185}, 2250 (2014)
  [arXiv:1310.1921 [hep-ph]].

\bibitem{Alwall:2014hca}
  J.~Alwall {\it et al.},
  JHEP {\bf 1407}, 079 (2014)
  [arXiv:1405.0301 [hep-ph]].

\bibitem{Ball:2014uwa}
  R.~D.~Ball {\it et al.} [NNPDF Collaboration],
  JHEP {\bf 1504}, 040 (2015)
  [arXiv:1410.8849 [hep-ph]].

\bibitem{Ade:2015xua}
  P.~A.~R.~Ade {\it et al.} [Planck Collaboration],
  Astron.\ Astrophys.\  {\bf 594}, A13 (2016)
  [arXiv:1502.01589 [astro-ph.CO]].

\bibitem{Anchordoqui:2011nh}
  L.~A.~Anchordoqui and H.~Goldberg,
  Phys.\ Rev.\ Lett.\  {\bf 108}, 081805 (2012)
  [arXiv:1111.7264 [hep-ph]].

\bibitem{Anchordoqui:2012qu}
  L.~A.~Anchordoqui, H.~Goldberg and G.~Steigman,
  Phys.\ Lett.\ B {\bf 718}, 1162 (2013)
  [arXiv:1211.0186 [hep-ph]].

\bibitem{Mangano:2001iu}
  G.~Mangano, G.~Miele, S.~Pastor and M.~Peloso,
  Phys.\ Lett.\ B {\bf 534}, 8 (2002)
  [astro-ph/0111408].

\bibitem{SolagurenBeascoa:2012cz}
  A.~Solaguren-Beascoa and M.~C.~Gonzalez-Garcia,
  Phys.\ Lett.\ B {\bf 719}, 121 (2013)
  [arXiv:1210.6350 [hep-ph]].

\bibitem{Kolb:1990vq}
  E.~W.~Kolb and M.~S.~Turner,
  Front.\ Phys.\  {\bf 69}, 1 (1990).

\bibitem{Amendola:2016saw}
  L.~Amendola {\it et al.},
  arXiv:1606.00180 [astro-ph.CO].

\bibitem{Belanger:2014vza}
  G.~Belanger, F.~Boudjema, A.~Pukhov and A.~Semenov,
  Comput.\ Phys.\ Commun.\  {\bf 192}, 322 (2015)
  [arXiv:1407.6129 [hep-ph]].
  G.~Belanger, F.~Boudjema, A.~Pukhov and A.~Semenov,
  Comput.\ Phys.\ Commun.\  {\bf 185}, 960 (2014)
  [arXiv:1305.0237 [hep-ph]].

\bibitem{Bertone:2004pz}
  G.~Bertone, D.~Hooper and J.~Silk,
  Phys.\ Rept.\  {\bf 405}, 279 (2005)
  [hep-ph/0404175].

\bibitem{Arcadi:2017kky}
  G.~Arcadi, M.~Dutra, P.~Ghosh, M.~Lindner, Y.~Mambrini, M.~Pierre, S.~Profumo and F.~S.~Queiroz,
  Eur.\ Phys.\ J.\ C {\bf 78}, no. 3, 203 (2018)
  [arXiv:1703.07364 [hep-ph]].

\bibitem{Akerib:2016vxi}
  D.~S.~Akerib {\it et al.} [LUX Collaboration],
  Phys.\ Rev.\ Lett.\  {\bf 118}, no. 2, 021303 (2017)
  [arXiv:1608.07648 [astro-ph.CO]].

\bibitem{Aprile:2017iyp}
  E.~Aprile {\it et al.} [XENON Collaboration],
  Phys.\ Rev.\ Lett.\  {\bf 119}, no. 18, 181301 (2017)
  [arXiv:1705.06655 [astro-ph.CO]].

\bibitem{Cui:2017nnn}
  X.~Cui {\it et al.} [PandaX-II Collaboration],
  Phys.\ Rev.\ Lett.\  {\bf 119}, no. 18, 181302 (2017)
  [arXiv:1708.06917 [astro-ph.CO]].

\bibitem{Aprile:2015uzo}
  E.~Aprile {\it et al.} [XENON Collaboration],
  JCAP {\bf 1604}, no. 04, 027 (2016)
  [arXiv:1512.07501 [physics.ins-det]].

\bibitem{DEramo:2016gos}
  F.~D'Eramo, B.~J.~Kavanagh and P.~Panci,
  JHEP {\bf 1608}, 111 (2016)
  [arXiv:1605.04917 [hep-ph]].

\bibitem{Pospelov:2007mp}
  M.~Pospelov, A.~Ritz and M.~B.~Voloshin,
  Phys.\ Lett.\ B {\bf 662}, 53 (2008)
  [arXiv:0711.4866 [hep-ph]].

\bibitem{Profumo:2017obk}
  S.~Profumo, F.~S.~Queiroz, J.~Silk and C.~Siqueira,
  JCAP {\bf 1803}, no. 03, 010 (2018)
  [arXiv:1711.03133 [hep-ph]].

\bibitem{Ackermann:2015zua}
  M.~Ackermann {\it et al.} [Fermi-LAT Collaboration],
  Phys.\ Rev.\ Lett.\  {\bf 115}, no. 23, 231301 (2015)
  [arXiv:1503.02641 [astro-ph.HE]].

\bibitem{Abdallah:2016ygi}
  H.~Abdallah {\it et al.} [H.E.S.S. Collaboration],
  Phys.\ Rev.\ Lett.\  {\bf 117}, no. 11, 111301 (2016)
  [arXiv:1607.08142 [astro-ph.HE]].

\bibitem{Abramowski:2011hc}
  A.~Abramowski {\it et al.} [H.E.S.S. Collaboration],
  Phys.\ Rev.\ Lett.\  {\bf 106}, 161301 (2011)
  [arXiv:1103.3266 [astro-ph.HE]].

\bibitem{Consortium:2010bc}
  M.~Actis {\it et al.} [CTA Consortium],
  Exper.\ Astron.\  {\bf 32}, 193 (2011)
  [arXiv:1008.3703 [astro-ph.IM]].

\bibitem{Lefranc:2015pza}
  V.~Lefranc, E.~Moulin, P.~Panci and J.~Silk,
  Phys.\ Rev.\ D {\bf 91}, no. 12, 122003 (2015)
  [arXiv:1502.05064 [astro-ph.HE]].

\bibitem{Rodejohann:2015lca}
  W.~Rodejohann and C.~E.~Yaguna,
  JCAP {\bf 1512}, no. 12, 032 (2015)
  [arXiv:1509.04036 [hep-ph]].

\bibitem{Griest:1990kh}
  K.~Griest and D.~Seckel,
  Phys.\ Rev.\ D {\bf 43}, 3191 (1991).

\bibitem{Gondolo:1990dk}
  P.~Gondolo and G.~Gelmini,
  Nucl.\ Phys.\ B {\bf 360}, 145 (1991).

\bibitem{Ibe:2008ye}
  M.~Ibe, H.~Murayama and T.~T.~Yanagida,
  Phys.\ Rev.\ D {\bf 79}, 095009 (2009)
  [arXiv:0812.0072 [hep-ph]].

\bibitem{Guo:2009aj}
  W.~L.~Guo and Y.~L.~Wu,
  Phys.\ Rev.\ D {\bf 79}, 055012 (2009)
  [arXiv:0901.1450 [hep-ph]].

\bibitem{Patra:2016ofq}
  S.~Patra, W.~Rodejohann and C.~E.~Yaguna,
  JHEP {\bf 1609}, 076 (2016)
  [arXiv:1607.04029 [hep-ph]].

\end{thebibliography}
\end{document}